\documentclass[orivec]{llncs}

\usepackage{natbib}
\usepackage[toc,page]{appendix}

\usepackage{aliascnt}
\usepackage{llncsdoc}
\usepackage{remreset}
\usepackage{authblk}
\newtheorem{Definition}{Definition}

\newcommand{\myappendices}[1]{%
\addcontentsline{app}{myappendices}{#1}\par}

\usepackage{lhelp}
\usepackage{amsfonts,latexsym}
\usepackage{amsmath}
\usepackage{amssymb}
\usepackage{caption}
\usepackage{algorithm}
\usepackage{algpseudocode}
\usepackage{MnSymbol}
\usepackage{verbatim}

\usepackage{listings}

\usepackage{hyperref}
\usepackage{cleveref}
\usepackage{color}

\usepackage{pgfplots}
\usepackage{tikz}
\usepackage{xlop}
\usetikzlibrary{positioning,calc}
\usetikzlibrary{arrows.meta,chains,decorations.pathreplacing,scopes,positioning}
\usetikzlibrary{calc,shapes.multipart,chains,arrows}
\graphicspath{ {figures/} }
\DeclareGraphicsExtensions{.eps,.png}

\usepackage{scrextend} 

\usepackage{hyperref}
\usepackage{lhelp}
\usepackage{listings}
\usepackage{float}
\usepackage{xcolor}
\newif\iflongversion
\longversionfalse

\def\Z{\mbox{${\mathbb Z}$}}

\def\0 {\ensuremath{\mathbf{0}}}

\algnewcommand\algorithmicswitch{\textbf{switch}}
\algnewcommand\algorithmiccase{\textbf{case}}
\algnewcommand\algorithmicassert{\texttt{assert}}
\algnewcommand\Assert[1]{\State \algorithmicassert(#1)}%
\algdef{SE}[SWITCH]{Switch}{EndSwitch}[1]{\algorithmicswitch\ #1\ \algorithmicdo}{\algorithmicend\ \algorithmicswitch}%
\algdef{SE}[CASE]{Case}{EndCase}[1]{\algorithmiccase\ #1}{\algorithmicend\ \algorithmiccase}%
\algtext*{EndSwitch}%
\algtext*{EndCase}%

\newcommand{\hidetext}[1]{\mbox{ \ }}

\newcommand{\textblockcomment}[1]{}
\renewcommand{\gets}{:=}

\newcommand{\MultTime}[1]{\mbox{{\sf M}$(#1)$}}
\def\DFT{\ensuremath{\mathrm{DFT}}}

\begin{document}

\title{Putting F{\"u}rer Algorithm into Practice with the BPAS Library}
\author{Sviatoslav Covanov\inst{2} \and Davood Mohajerani  \and  Marc~Moreno~Maza\inst{4} \and Lin-Xiao Wang\inst{5} \\
Department of Computer Science, The University of Western Ontario}
\institute{
  \email{svyatoslav.covanov@gmail.com},
  \and
  \email{mohajerani.d@gmail.com},
  \and
  \email{moreno@csd.uwo.ca},
  \and
  \email{lwang739@uwo.ca}.
}

\maketitle

\begin{abstract}

Fast algorithms for integer and polynomial multiplication play an
important role in scientific computing as well as in other disciplines.
In 1971, Sch{\"o}nhage and Strassen designed an algorithm that
improved the multiplication time for two integers of at most $n$ bits to
$\mathcal{O}(\log n \log \log n)$. In 2007, Martin F{\"u}rer presented
a new algorithm that runs in $O \left(n \log n\ \cdot 2^{O(\log^* n)}
\right)$, where $\log^* n$ is the iterated logarithm of $n$.

We explain how we can put F{\"u}rer's ideas into practice for
multiplying polynomials over a prime field $\mathbb{Z} / p
\mathbb{Z}$, for which $p$  is a Generalized Fermat prime of the
form $p = r^k + 1$ where $k$ is a power of $2$ and $r$ is of machine
word size.  When $k$ is at least 8, we show that multiplication inside
such a prime field can be efficiently implemented via Fast Fourier
Transform (FFT).  Taking advantage of Cooley-Tukey tensor formula and the
fact that $r$ is a $2k$-th primitive root of unity in $\mathbb{Z} / p
\mathbb{Z}$, we obtain an
efficient implementation of FFT over $\mathbb{Z} / p \mathbb{Z}$.
This implementation outperforms comparable implementations either
using other encodings of $\mathbb{Z} / p \mathbb{Z}$ or other ways to
perform multiplication in $\mathbb{Z} / p \mathbb{Z}$.

\end{abstract}

\section{Introduction}
\label{sec:Introduction}

Asymptotically fast algorithms for exact polynomial and matrix
arithmetic play a central role in scientific computing.  Among others,
the discoveries of Karatsuba~\cite{KO63}, Cooley and
Tukey~\cite{CooleyTukey}, Strassen~\cite{Strassen69}, and 
Sch{\"{o}}nhage and Strassen~\cite{DBLP:journals/computing/SchonhageS71}
have
initiated intense activity in both numerical and symbolic
computing.  The implementation of asymptotically fast algorithms is a
 research direction in its own right. Theoretical analyses of
asymptotically fast algorithms typically focus on arithmetic operation counts,
thereby ignoring important hardware details, in particular the
costs of memory accesses.
On modern hardware architectures such theoretical simplifications
are of questionable value, and other complexity measures, such as cache
complexity~\cite{DBLP:journals/talg/FrigoLPR12}, are needed
to better analyze algorithms.

The algorithm of Sch{\"{o}}nhage and
Strassen~\cite{DBLP:journals/computing/SchonhageS71} is an
asymptotically fast algorithm for multiplying integers in arbitrary
precision.  It uses the fast Fourier transform (FFT) and, for two
integers of at most $n$ bits, it computes their product in $O (n {\log}
n \cdot {\log} {\log} n)$ bit operations\footnote{We denote by ${\log}(n)$
the logarithm of the positive integer $n$ with respect to base $2$
and, for a positive real number $b$, we denote by ${\log}_b(n)$
the logarithm of $n$ with respect to base $b$}.  This result remained the best
known upper bound until the celebrated paper of Martin
{F\"{u}rer}~\cite{DBLP:conf/stoc/Furer07}.  His integer multiplication
algorithm runs in $O \left(n {\log} n\ \cdot 2^{O(\log^* n)} \right)$
bit operations, where $\log^* n$ is the iterated logarithm of $n$,
defined as:
\begin{equation}
 \log^* n :=
  \begin{cases}
    0                  & \mbox{if } n \le 1; \\
    1 + \log^*(\log n) & \mbox{if } n > 1
   \end{cases}
\end{equation}
A detailed analysis suggests that {F\"{u}rer's} algorithm
is expected to outperform
that of Sch{\"{o}}nhage and Strassen for $n \geq 2^{2^{64}}$.

The practicality of {F\"{u}rer's} algorithm is still an open
question, a question that we address in this paper.
Before presenting our approach, we observe that the
ideas of F\"{u}rer are not specific to integer multiplication
and can be used for multiplying polynomials with coefficients
in the field $\mathbb{C}$ of complex numbers or in any finite field.
In~\cite{DBLP:conf/stoc/DeKSS08,DBLP:journals/siamcomp/DeKSS13}
De et al.{}
gave a similar algorithm which relies on {\em finite field arithmetic} 
and achieves the same running time as {F\"{u}rer's} algorithm.
We adopt such a framework here, taking polynomials to have coefficients
in a finite field.

\subsection{The ``main trick'' of F\"{u}rer's algorithm}
For this exposition
we follow an analysis reported by Chen et al.{}
in~\cite{DBLP:conf/issac/ChenCMM17}. Consider a prime field ${\Z}/p{\Z}$ and $N$, a power of $2$,
dividing $p-1$.
Then, the finite field ${\Z}/p{\Z}$  admits an $N$-th primitive root 
of unity.
Denote such an element by ${\omega}$.
Let $f \in {\Z}/p{\Z}[x]$ be a polynomial  of degree at most $N-1$.
Then, computing the discrete Fourier transform (DFT) of $f$ at ${\omega}$
produces
the values of $f$ at successive powers of  ${\omega}$,
that is, $f({\omega}^0), f({\omega}^1), \ldots f({\omega}^{N-1})$.
Using an asymptotically fast algorithm, namely a fast Fourier
transform (FFT), this calculation 
amounts to:
\begin{enumerateshort}
\item $N \, \log(N)$ additions in ${\Z}/p{\Z}$; and
\item $(N/2) \, \log(N)$ multiplications by a power of ${\omega}$ in ${\Z}/p{\Z}$.
\end{enumerateshort}
If the size of $p$ is $k$ machine words, then 
\begin{enumerateshort}
\item each addition in ${\Z}/p{\Z}$ costs $O(k)$ machine-word operations; and
\item each multiplication by a power of ${\omega}$ costs
      $O( \MultTime{k}  )$ machine-word operations,
\end{enumerateshort}
where $n \longmapsto \MultTime{n}$ is a multiplication time
as defined in \cite{Gathen:2003:MCA:945759}
Therefore, multiplication by a power of ${\omega}$
becomes a bottleneck as $k$ grows.

To overcome this difficulty, we consider the following trick
proposed by Martin F\"{u}rer 
in~\cite{DBLP:conf/stoc/Furer07,DBLP:journals/siamcomp/Furer09}.
We assume that $N = K^e$ holds for some ``small'' $K$, say $K = 32$
and an integer $e \geq 2$.
Further, we define ${\eta} = {\omega}^{N/K}$ and $J = K^{e-1}$ and assume that 
multiplying an arbitrary element of ${\Z}/p{\Z}$
by ${\eta}^i$, for any $i = 0, \ldots, K-1$, can be done within
$O(k)$ machine-word operations. 
Consequently, every arithmetic operation (addition, multiplication) involved in 
a DFT on $K$ points, using ${\eta}$ as a primitive root, amounts to 
$O(k)$ machine-word operations. 
Therefore, a DFT of size $K$ can be performed 
with $O(K \, \log(K) \, k)$ machine-word operations, rather than
$O(K \, \log(K) \, \MultTime{k})$ without the assumed special value of $N$.
Since the multiplication time  $n \longmapsto \MultTime{n}$ is necessarily
super-linear, the former estimate is asymptotically smaller
than the latter one.
As we shall see in Section~\ref{sec:GFPF},
this result holds 
whenever $p$ is a so-called {\em generalized Fermat number}.

Returning to the DFT of size $N$ at ${\omega}$ and 
using the factorization formula of Cooley and Tukey~\cite{CooleyTukey}, we have 
\begin{equation} 
\label{eq:CTCT}
\DFT_{J K} \ = \ (\DFT_{J} \otimes I_K) D_{J,K} (I_J \otimes \DFT_{K}) L_J^{J K},
\end{equation}
where the elements of this equation are as defined in Section~\ref{sec:FFT}.
Hence, the DFT of $f$ at ${\omega}$ is essentially performed by:
\begin{enumerateshort}
\item $K^{e-1}$ DFT's of size $K$ (that is, DFT's on polynomials
      of degree at most $K-1$),
\item $N$ multiplications by a power of ${\omega}$ (coming from the diagonal matrix
       $D_{J,K}$) and 
\item $K$ DFT's of size $K^{e-1}$.
\end{enumerateshort}
Unrolling Formula (\ref{eq:CTCT}) so as to replace $\DFT_{J}$
by $\DFT_{K}$ and the other linear operators involved
(the diagonal matrix $D$ and the permutation matrix $L$)
one can see that a DFT of size $N = K^e$ reduces to:
\begin{enumerateshort}
\item $e \, K^{e-1}$ DFT's of size $K$, and
\item $(e-1) \, N$ multiplications by a power of ${\omega}$.
\end{enumerateshort}
Recall that the assumption on the cost of a multiplication
by ${\eta}^i$, for $0 \leq i < K$, makes the  
cost for one DFT of size $K$ to $O(K \, \log_2(K) \, k)$ machine-word
operations.
Hence, all the DFT's of size $K$ together
amount to $O(e \, N \, \log_2(K) k)$ machine-word
operations. That is, $O(N \, \log_2(N) \, k)$ machine-word
operations.
Meanwhile, the total cost of the 
multiplication by a power of ${\omega}$ is
$O(e \, N \,  \MultTime{k} )$ machine-word
operations, that is, $O(N \, \log_{K}(N) \,  \MultTime{k} )$ machine-word
operations.
Indeed, multiplying an arbitrary element of ${\Z}/p{\Z}$ by 
an arbitrary power of ${\omega}$ requires $O(  \MultTime{k} )$ machine-word
operations.
Therefore, under our assumption, a DFT of size $N$ at ${\omega}$ 
amounts to
\begin{equation}
\label{eq:BigPrimeFieldFFTTootalCost}
O(N \, \log_2(N) \, k \ + \ N \, \log_{K}(N) \, \MultTime{k})
\end{equation}
machine-word operations.
When using generalized Fermat primes, we have $K = 2 k$
and the above estimate becomes
\begin{equation}
\label{eq:BigPrimeFieldFFTTootalCostTwo}
O(N \, \log_2(N) \, k \ + \ N \, \log_{k}(N) \, \MultTime{k}).
\end{equation}
The second term in the big-O notation dominates
the first one. 
Without our assumption, as discussed earlier, the same DFT would
run in $O(N \, \log_{2}(N) \, \MultTime{k})$ machine-word
operations. Therefore, using generalized Fermat primes
brings a speedup factor of $\log(K)$ 
w.r.t. the direct approach using arbitrary prime 
numbers.

\subsection{Overview}
We are addressing two questions in this paper.  First, can we observe
the above described speedup factor on a serial implementation written
in the programming language C and run on modern multicore processors?
Indeed, the authors of~\cite{DBLP:conf/issac/ChenCMM17} answered a
similar question in the case of a CUDA implementation targeting GPUs
(Graphics Processing Units).  Such architectures offer to programmers
a finer control of hardware resources than multicore processors, thus
more opportunities to reach high performance.  Hence, this first
question is a natural challenge.

Second, can we use FFT to implement multiplication
in ${\Z}/p{\Z}$ and obtain better performance
than using plain multiplication in ${\Z}/p{\Z}$
This was not attempted in the GPU implementation
of~\cite{DBLP:conf/issac/ChenCMM17}.
However this is a natural question in the 
spirit of the algorithms 
of Sch{\"{o}}nhage and Strassen~\cite{DBLP:journals/computing/SchonhageS71}
and {F\"{u}rer}~\cite{DBLP:conf/stoc/Furer07}, where
fast multiplication is achieved by ``composing'' 
FFTs operating on different vector sizes.
The experimental results reported in Section~\ref{sec:Experimentation}
give positive answers to both questions.

Consider a Generalized Fermat prime number of the form $p = r^k + 1$, 
where $k$ is a power of 2 and $r$ is of machine-word size.
As mentioned above, as well as in~\cite{DBLP:conf/issac/ChenCMM17},
multiplying by a power of $r$ modulo $p$ can be in $O(k)$
machine-word operations. However, multiplying two arbitrary
elements of $\mathbb{Z} / p \mathbb{Z}$ is a non-trivial
operation. Note that we encode elements of $\mathbb{Z} / p \mathbb{Z}$
in radix $r$ expansion. Thus, multiplying two arbitrary
elements of $\mathbb{Z} / p \mathbb{Z}$ requires
to compute the product of two univariate polynomials 
in ${\Z}[X]$, of degree less than $k$, modulo $X^k + 1$.
In~\cite{DBLP:conf/issac/ChenCMM17}, this is done
by using plain multiplication, thus ${\Theta}(k^2)$
machine-word operations.
In Section~\ref{sec:GFPF}, we explain how to multiply
two arbitrary
elements $x, y$ of $\mathbb{Z} / p \mathbb{Z}$ via FFT.
We give a detailed analysis of the algebraic complexity
of our procedure.
A natural alternative to our approach would be
to compute $(x y) \mod{p}$ where the product $x y$ is 
an integer computed after converting the radix $r$ expansion
of $x, y$ to integers (say in binary expansions).
We show that this alternative approach is theoretically
and practically less efficient than the one via FFT.

To verify the benefits
of F\"{u}rer's trick experimentally we need to perform
FFT computations over a Generalized Fermat
prime field ${\Z}/p{\Z}$, for different implementations
of that prime field, as follows.
Either one should be able to assume that the elements of ${\Z}/p{\Z}$
are in radix $r$ expansion (when $p$ writes $r^k +1$
where $k$ is a power of $2$) or one should
simply be able to use traditional radix $2$ expansions.
Moreover, we consider multiplying two arbitrary
elements of  ${\Z}/p{\Z}$ via FFT.
Overall, we need an  implementation
of FFT running over a variety of prime fields.
Section~\ref{sec:FFT} reports on a
generic implementation of FFT over
finite fields in the BPAS library~\cite{DBLP:conf/icms/ChenCMMXX14}.

\section{Generalized Fermat prime fields}
\label{sec:GFPF}
A {\em Galois field}, also known as {\em finite field}, 
is a field with finitely many elements.
The residue classes modulo $p$, 
where $p$ is a prime number, form a field 
(unique up to isomorphism)
called the  {\em prime field} with $p$ elements
and denoted by $\mathsf{GF}(p)$ or $\mathbb{Z} / p \mathbb{Z}$.
Single-precision and multi-precision primes are referred to as
{\em small primes} and {\em big primes}, respectively.

Arithmetic operations for polynomials and matrices over prime fields
play a central role in computer algebra. Efficient implementations of 
these operations support the computation
over Galois fields that are essential to cryptography algorithms as
and coding theory. In symbolic computation, 
the implementation of modular methods,
prime fields are often using machine word
size characteristic. Increasing the arithmetic to
greater precision can be done using the Chinese Remainder Theorem
(CRT).

However, using these small prime numbers can cause problems in some certain modular methods. In particular, we must avoid choosing \textit{unlucky primes}, where the correct output is one of several quantities with the same modular image~\cite{Arnold:2003:MAC:937627.937629,DBLP:conf/issac/DahanMSWX05}. Because of the limitations of using small prime numbers, arithmetic over prime fields for multi-precision primes is desirable for some problems, such as polynomial system solving~\cite{DBLP:conf/issac/DahanMSWX05}.

Since modular methods for polynomial systems rely on polynomial
arithmetic, those large prime numbers must support FFT-based
algorithms, such as FFT-based polynomial multiplication.  This leads
us to consider the so-called Generalized Fermat prime numbers.

The $n$-th Fermat number can be denoted by $F_n = 2^{2^n} + 1$. This
sequence of numbers plays an essential role in number
theory. Arithmetic operations on fields based on Fermat numbers are
simpler than those of other arbitrary prime numbers since 2 is the
$2^{n+1}$-th primitive root of unity modulo $F_n$. But, unfortunately,
the largest Fermat prime number known now is $F_4$. This triggered the
interests of finding Fermat-like numbers. \textit{Generalized Fermat
	numbers} are one of these kinds.

Numbers that are in the form of ${a^{2^{n}}} + b^{2^{{n}}}$ with $a$,
$b$ any co-prime integers, where $a > b > 0$ and $n > 0$ hold, are
called \textit{generalized Fermat numbers}. Among all, those with $b =
1$ are of the most interest; we commonly write generalized Fermat
numbers of the form $a^{2^{n}} + 1$ as $F_n(a)$. For a prime generalized
Fermat number $p$, we use $\mathbb{Z}/p\mathbb{Z}$ to represent the
finite field $\mathsf{GF}(p)$. In particular, in the field
$\mathbb{Z}/F_n(a)\mathbb{Z}$, $a$ is a $2^{n+1}$-th primitive root
of unity. But with the binary representation of numbers on computers,
the arithmetic operations on such fields are not as simple as those of
Fermat numbers. To solve this problem, a special kind of generalized
Fermat number is defined in the previous work of our research
group~\cite{DBLP:conf/issac/ChenCMM17}.

Any integer in the form of $F_n(r) = (2^w \pm 2^u)^k + 1$ is called a sparse radix generalized Fermat number, where $w > u \ge 0$. Table \ref{tab:SparseRadixGeneralizedFermatNumbers} lists some sparse radix generalized Fermat numbers that are primes. For each prime $p = F_n(r)$, $k$ is some power of 2 and the prime writes as $p = r^k + 1$. In the same table, the number 
$2^e$ is the largest power of 2 that divides $p - 1$, which gives the maximum length of a vector to which we can apply a 2-way FFT algorithm.

\begin{table}[htbp]
	\centering%
	\caption{\small SRGFNs of practical interest.}
	\renewcommand*{\arraystretch}{1.35}
	{
		\begin{scriptsize}
			\begin{tabular}{|l|l|} \hline 
				$p$ & ${\max}\{ 2^e \ {\rm s.t.} \ 2^e \, \mid \,  p-1 \}$ \\ \hline  \hline
				$(2^{63} + 2^{53})^2 + 1$ & $2^{106}$ \\ \hline 
				$(2^{64} - 2^{50})^4 + 1$ & $2^{200}$ \\ \hline 
				$(2^{63} + 2^{34})^8 + 1$ & $2^{272}$ \\ \hline 
				$(2^{62} + 2^{36})^{16} + 1$ & $2^{576}$ \\ \hline  
				$(2^{62} + 2^{56})^{32} + 1$ & $2^{1792}$ \\ \hline 
				$(2^{63} - 2^{40})^{64} + 1$ & $2^{2560}$ \\ \hline 
				$(2^{64} - 2^{28})^{128} + 1$ & $2^{3584}$ \\ \hline 
			\end{tabular}
		\end{scriptsize}
	}
	\label{tab:SparseRadixGeneralizedFermatNumbers}
\end{table}

In the finite prime field $\mathbb{Z}/p\mathbb{Z}$, where $p = r^k +1$, each element $x$ is represented by a vector $\vec{x} = (x_{k-1},\ldots, x_0)$ of length $k$. We restrict all the coefficients to be non-negative integers so that we have 
\begin{eqnarray}
X \equiv x_{k-1}\,r^{k-1} + x_{k-2}\,r^{k-2} + \cdots + x_1\,r + x_0 \mod p \label{ElementRepresentation}
\end{eqnarray}
The following two cases make the representation unique for each element:
\begin{enumerate}
	\item When $x \equiv p - 1 \mod p$ holds, we have $x_{k-1} = r$ and $x_{k-2} = \cdots = x_0 = 0$.
	\item When $0 \le x <  p - 1$ holds, we have $0 \le x_i < r$ for $i = 0,\ldots,k-1$.
\end{enumerate}
We can also use a univariate polynomial $f_x \in \mathbb{Z}[R]$ to
represent $x$: 
we write $f_x = \sum_{i = 0}^{k-1} x_i\,R^i$, such that $x \equiv f_x(r) \mod p$.

\subsection{Computing the primitive root of unity in $\mathbb{Z}/p\mathbb{Z}$}
\label{sec:pritimiverootingfpf}
Primitive roots of unity
are special elements in a field
that are required by some algorithms, such as  Fast
Fourier transforms. For a field
$\mathbb{F}$ and an integer $n \ge 1$, an element
$\omega \in \mathbb{F}$ is an 
{\em $n$-th primitive root of unity}, if it meets the
following two requirements~\cite{MCA}.

\begin{enumerate}
	\item[$(i)$] $\omega$ is an {\em $n$-th root of unity}, that is, 
	we have $\omega ^ n = 1$.
	\item[$(ii)$] we have $\omega^{i} \neq 1$ for all $1 < i < n$.
\end{enumerate}

For any $n$ that divides $p - 1$,there are various ways to find an $n$-th primitive root of unity in
$\mathbb{Z}/p\mathbb{Z}$. Now we want to consider the case of finding
an $N$-th primitive root of unity $\omega$ in $\mathbb{Z}/p\mathbb{Z}$
such that ${\omega}^{N/2k} = r$ holds. Indeed, computing 
a DFT at such $\omega$ on a vector of size $N$ would take advantage of the fact
that multiplying by a power of $r$ can be done in linear time (see Section~\ref{sec:multiplybypowerofr}).

In Algorithm~\ref{algo:GFPFRootOfUnity}, the input 
$N$ is a power of 2 that divides $p - 1$ 
and the input $g$ is a $N$-th primitive root of unity in $\mathbb{Z}/p\mathbb{Z}$.

\begin{algorithm}[h]
	\caption{Primitive $N$-th root ${\omega} \in {\mathbb{Z}}/p{\mathbb{Z}}$ such that ${\omega}^{N/2 k} = r$ }
	\label{algo:GFPFRootOfUnity}
	\begin{algorithmic}[1]
		\Procedure{BigPrimeFieldPrimitiveRootOfUnity}{$N, r, k, g$}
		\State $a := g^{{N/2k}}$
		\State $b := a$
		\State $j := 1$
		\While{$b \neq r$}
		\State $b := a\,b$
		\State $j := j + 1$
		\EndWhile
		\State ${\omega} := g^j$
		\State \Return $({\omega})$
		\EndProcedure
	\end{algorithmic}
\end{algorithm}

From the definition of generalized Fermat prime numbers we know that $r$ is a $2k$-th primitive root of unity in $\mathbb{Z}/p\mathbb{Z}$, where $p = r^k + 1$. While $g^{N/2k}$ is a $2k$-th root of unity, it must equal to some power of $r$, say $r^t\mod p$ for some $0 \le t < 2k$. Let $j$ be a non-negative integer, $q$ and $s$ are the quotient and the remainder of $j$ in the Euclidean division by $2k$, so we have
\begin{eqnarray}
j = q \cdot 2k + s
\end{eqnarray}
and
\begin{eqnarray}
g^{jN/2k} = g^{2kq+s}\,g^{N/2k} = g^s\,g^{N/2k} = (g^{N/2k})^s = r^{ts}
\end{eqnarray}

By the definition of primitive root of unity, the powers $r^{ts}$ are pairwise different for $0 \le s < 2k$ and for some $s_i$, $r^{ts_i} = r$ holds. Hence, for some $j_i = q_i \cdot 2k + s_i$, we will have $(g^{N/2k})^{j_i} = r$. Then $\omega = g^{j_i}$ is the primitive root of unity that we want.

\subsection{Addition and subtraction in $\mathbb{Z}/p\mathbb{Z}$}
\label{sec:addandsubingfpf}
Let $x,y \in \mathbb{Z}/p\mathbb{Z}$ represented by vectors $\vec{x}$ and $\vec{y}$. The following algorithm \ref{algo:BigPrimeFieldAddition} computes $\overrightarrow{x+y}$ that represents the sum of $x$ and $y$ in $\mathbb{Z}/p\mathbb{Z}$. We firstly compute the component-wise addition of $\vec{x}$ and $\vec{y}$ with carry. If there's no carry beyond the last component $u_{k-1}$, then ${u_0,\ldots,u_{k-1}}$ is the vector representation of $x + y$ in $\mathbb{Z}/p\mathbb{Z}$. If there is a carry, then the sum is over $r^k$ and we need to do a subtraction of carry by the vector $\vec{u}$, since $r^k \equiv p - 1 \equiv -1 \mod p$.

\begin{algorithm}[H]
	\caption{Computing 
		$x + y \in {\mathbb{Z}}/p{\mathbb{Z}}$ for 
		$x, y \in {\mathbb{Z}}/p{\mathbb{Z}}$}
	\label{algo:BigPrimeFieldAddition}
	\begin{algorithmic}[1]
		\Procedure{BigPrimeFieldAddition}{$\vec{x}, \vec{y}, r, k$}
		\State compute $z_i = x_i + y_i$ in $\mathbb{Z}/p\mathbb{Z}$, for $i = 0, \ldots, k-1$,
		\State let $z_k = 0$,
		\State for $i = 0, \ldots, k-1$, compute the quotient $q_i$
		and the remainder $s_i$ in the Euclidean division of
		$z_i$ by $r$, then replace $(z_{i+1}, z_i)$
		by $(z_{i+1} + q_i, s_i)$,
		\State if $z_k = 0$ then return $(z_{k-1}, \ldots, z_0)$,
		\State if $z_k = 1$ and $z_{k-1} = \cdots = z_0 = 0$,
		then let $z_{k-1} = r$ and return $(z_{k-1}, \ldots, z_0)$,
		\State let $i_0$ be the smallest index, $0 \leq i_0 \leq k$,
		such that $z_{i_0} \neq 0$, then let $z_{i_0} = z_{i_0} - 1$, let
		$z_0 = \cdots  = z_{i_0 - 1} = r-1$ and return $(z_{k-1}, \ldots, z_0)$.
		\EndProcedure
	\end{algorithmic}
\end{algorithm}

In this theoretical algorithm, we use a Euclidean division to compute the carry and the remainder of $x_i + y_i$, which requires a division and a subtraction operation. But in practical implementation, we can avoid the expensive division. The following lists the C code we used in the BPAS library.

According to the method above, each component of $\vec{x}$ and $\vec{y}$ is in the range of $[0,k-1]$ for $0 \le i \le k - 2$ and $x_{k-1},y_{k-1} \in [0,k]$, so that we can safely say that the results of the component-wise addition will not be greater than $2r - 2$ for the first $k-1$ pairs of component. Hence, if the sum is greater than $r$, we can simply subtract the result by $r$ and set the carry to 1, instead of using an Euclidean division. For the last pair $x_{k-1}$ and $y_{k-1}$, the two special cases are one of them is equal to $r$ and both of them are equal to $r$. For the first case, the maximum sum of $x_{k-1}$ and $y_{k-1}$ is $2r-1$, there is no difference from the previous method. 

Now, let's consider the second case where both $x_{k-1}$ and $y_{k-1}$ are equal to $r$. And all of the other components in the vectors are 0, such that both $x$ and $y$ are equal to $r^k$. When we add the two components together $u_{k-1}$ is equal to $2r$ and by using line 9 to 11 from listing \ref{lst:AdditionInGeneralizedFermatprimefield}, we have $u_{k-1} = r$ and carry = 1. Then $u_{k-1}$ is the first $u_i$ that is not 0. In line 33, we have $u_{k-1} = u_{k-1} - 1 = r - 1$ and in line 31 we set $u_i = r - 1$ for $0 \le i < k - 1$. The result we get is $u_i = r - 1$ for $0 \le i < k$, that is equal to $u \equiv -2 \mod p$, indeed that $x + y \equiv r^k + r^k \equiv 2(p - 1) \equiv 2p - 2 \equiv -2 \mod p$. So far we have proved that our algorithm works correctly and efficiently for all of the cases.

\begin{lstlisting}[caption={Addition in a Generalized Fermat Prime Field},captionpos=b,label = {lst:AdditionInGeneralizedFermatprimefield}]
sfixn* addition (sfixn * x, sfixn *y, int k, sfixn r) {
	short c = 0;
	short post = 0;
	sfixn sum = 0;
	int i = 0;

	for (i=0;i < k;i++) {
		sum = x[i] + y[i] + c;
		if (sum >= r ) {
			c = 1;
			x[i] = sum - r;
		}
		else {
			x[i] = sum;
			c = 0;
		}
	}

	if (c > 0){
		post = -1;
		for (i = 0; i < k; i++) {
			if (x[i] != 0){
			post = i;
			break;
			}
		}

		if (post >= 0){
			for (i = 0; i < post; i++) {
				x[i] = r - 1;
			}
			x[post]--;
		}
		else {
			x[k-1] = r;
			for (i = 0;i < k-1; i++){
				x[i] = 0;
			}
		}
	}	

	return x;
}
\end{lstlisting}

Similarly, we have an algorithm $\mathrm{BigPrimeFieldSubtraction} (\vec{x}, \vec{y}, r, k)$ for computing $\overleftrightarrow{x - y}$ represents $(x - y) \in \mathbb{Z}/p\mathbb{Z}$.

\subsection{Multiplication by power of $r$ in $\mathbb{Z}/p\mathbb{Z}$}
\label{sec:multiplybypowerofr}
Multiplication between two arbitrary elements in $\mathbb{Z}/p\mathbb{Z}$ can be very complicated and expensive, and Chapter 4 will explain that process in greater detail. Now, let us consider the case of multiplication between elements $x,y \in \mathbb{Z}/p\mathbb{Z}$, where one of them is a power of $r$. We assume that $y = r^i$ for some $0 \le i \le 2k$. The cases that $i = 0$ and $i = 2k$ are trivial, since $r$ is a $2k$-th primitive root of unity in $\mathbb{Z}/p\mathbb{Z}$, we have $r^0 = r^{2k} = 1$. Also we have $r^k = -1$ in $\mathbb{Z}/p\mathbb{Z}$, so that for $i = k$, we have $x = -x$ and for $k < i < 2k$, $r^i = -r^{i - k}$ holds. Now let us only consider the case that $0 < i < k$, where we have the following equation:
\begin{equation*}
\begin{array}{rcl}
x r^i & \equiv & (x_{k-1} \, r^{k-1+i} + \cdots + x_0 \, r^i )\ \mod{p}  \\[1.5ex]
& \equiv & \sum\limits_{j=0}^{j=k-1} \ x_j r^{j+i} \ \mod{p} \\
& \equiv & \sum\limits_{h = i}^{h = k-1+i} \ x_{h-i} r^h    \ \mod{p} \\[1.6ex]
& \equiv & (\sum\limits_{h = i}^{h = k-1} \ x_{h-i} r^h  -
\sum\limits_{h = k}^{h = k-1+i} \ x_{h-i} r^{h-k})  \ \mod{p} \\[1.6ex]
\end{array}
\end{equation*}

We see that for all $0 \le i \le 2k$, $x\cdot r^i$ is reduced to some shift and a subtraction. We call this process cyclic shift. The following gives the C implementation in the BPAS library.

\begin{lstlisting}[caption={Multiplication by a power of $r$ in a Generalized Fermat Prime Field},captionpos=b,label = {lst:MulPowRGFPF}]
sfixn* MulPowR(sfixn *x,int s, int k, sfixn r){
	sfixn *a =(sfixn*)calloc(sizeof(sfixn),k);
	sfixn *b =(sfixn*)calloc(sizeof(sfixn),k);
	sfixn *c =(sfixn*)calloc(sizeof(sfixn),k);
	s = s%(2 * k);
	if (s == 0)
		return x;
	else if (s == k)
		return BigPrimeFieldSubtraction(c,x,k,r);
	else if ((s > k) && (s < (2 * k))){
		s = s - k;
		x = BigPrimeFieldSubtraction(c,x,k,r);
	}
	int i;
	for (i = 0; i < (k - s); i++)
		b[i + s] = x[i];
	for (i = k - s; i < k; i++)
		a[i - (k - s)] = x[i];
	if(x[k-1] == r){
		a[s-1] -=r;
		a[s] ++;
	}
	return BigPrimeFieldSubtraction(b,a,k,r);
}
\end{lstlisting}

\subsection{Multiplication between arbitrary elements in $\mathbb{Z}/p\mathbb{Z}$}
\label{sec:bigprimefieldmultiplication}

According to previous explanation, we use the univariate
polynomials $f_x,f_y \in \mathbb{Z}[R]$ to represent input elements $x,y \in
\mathbb{Z}/p\mathbb{Z}$ respectively. 
Algorithm~\ref{algo:BigPrimeFieldMultiplication} then computes the product $x\cdot y
\in \mathbb{Z}/p\mathbb{Z}$. In the first step, we multiply the two
polynomials over $\mathbb{Z}$ and compute the remainder $f_u$ 
of the product modulo
$R^k + 1$. Then, we convert all the coefficients of $f_u$ into the
radix-$r$ representation in $\mathbb{Z}/p\mathbb{Z}$. Finally we
multiply each coefficient with the corresponding power of $r$ using the
``cyclic shift'' operation from Section~\ref{sec:multiplybypowerofr} and add all the
results together.

\begin{algorithm}[H]
	\caption{Computing 
		$x \cdot y \in {\mathbb{Z}}/p{\mathbb{Z}}$ for 
		$x,y \in {\mathbb{Z}}/p{\mathbb{Z}}$}
	\label{algo:BigPrimeFieldMultiplication}
	\begin{algorithmic}[1]
		\State {\textbf{input:} \begin{itemize}
				\item[-] an integer $\tt k$ and radix $\tt r$,
				\item[-] two polynomials $f_x$ and $f_y$ whose coefficient vectors are $\vec{x}, \vec{y}$.
			\end{itemize}
		}
		\State {\textbf{output:} \begin{itemize}
				\item[-] a vector $\vec{u}$
			\end{itemize}
		}
		\Procedure{BigPrimeFieldMultiplication}{$f_x, f_y, r, k$}
		\State $f_u(R) := f_x(R) \cdot f_y(R)$
		\Comment computing $f_x$ times $f_y$ in $\mathbb{Z}[R]$
		\State $f_u(R) := f_u(R) \mod (R^k + 1)$
		\Comment we get $f_u(R) = \sum_{i = 0}^{k - 1}u_i \cdot R^i$
		\State $\vec{u}$ is the coefficient vector of $f_u$
		\For{$0 \le i < k$}
		\State $\vec{u_i} := u_i \in \mathbb{Z}/p\mathbb{Z}$
		\Comment compute a radix representation of each $u_i$
		\EndFor
		\State $\vec{u} := \vec{u_0}$
		\Comment add all the $u_i$ together using the algorithm \ref{algo:BigPrimeFieldAddition}
		\For{$1 \le i < k$}
		\State $\vec{u}$ :=BigPrimeFieldAddition$(\vec{u},\vec{u_i},k,r)$
		\EndFor
		\State return $\vec{u}$
		\EndProcedure
	\end{algorithmic}
\end{algorithm}

In the following section, we will discuss the multiplication between arbitrary elements in more detail, and we analyze the different implementations of the algorithm.

\section{Optimizing multiplication in Generalized Fermat prime fields}
\label{GFPFMulti}

In this section, we will discuss how to multiply two arbitrary elements
in $\mathbb{Z}/p\mathbb{Z}$ efficiently using FFT,
when $p$ is a Generalized Fermat prime.
Firstly, in Section~\ref{sec:algorithmformultiplication}, we outline
two algorithms that we can use for this multiplication: one is based
on polynomial multiplication (see Section~\ref{sec:fft-based}) and the
other one is based on integer multiplication by means of the GMP
library~\cite{Granlund12} (see
Section~\ref{sec:integermultiplication}).
Then, in Section~\ref{sec:analysisofmultiplication}
we provide detailed complexity analysis on the two
approaches. Finally, in Section~\ref{sec:fft-based implementation}, we
present the implementation of the FFT-based polynomial-based
multiplication. We break
down the algorithm into sub-routines and explain in details for each part. 
The C functions that we use can be found in Appendix~\ref{AppA}.
 
\subsection{Algorithms}
\label{sec:algorithmformultiplication}
Let $p$ be a Generalized Fermat prime.
When actually implementing the multiplication of two arbitrary
elements in the field $\mathbb{Z}/p\mathbb{Z}$, we use two different
approaches. In the first approach, we follow the basic idea explained
in Section~\ref{sec:GFPF} (see Algorithm~\ref{algo:BigPrimeFieldMultiplication}) which
treats any two elements $x, y$ in the field as polynomials 
$f_x, f_y$ and uses polynomial
multiplication algorithms to compute the product $x y$. The other approach
involves converting the elements $x, y$ from their radix-$r$ representation
into GMP integer numbers and letting the GMP library~\cite{Granlund12}
do the job.

\subsubsection{Modular multiplication based on polynomial multiplication}
\label{sec:fft-based}

In Section~\ref{sec:bigprimefieldmultiplication} we gave the basic
algorithm for multiplying two arbitrary elements of
$\mathbb{Z}/p\mathbb{Z}$ based on polynomial multiplication.  In
practice, there are more details to be considered in order to reach
high-performance.  For instance, how do we efficiently convert a
positive integer in the range $(o, r^3)$ into radix-$r$
representation.

Let us consider how to calculate $u = x \, y \mod p$ with $x,y,u \in
\mathbb{Z}/p\mathbb{Z}$. Here we want to use the polynomial
representation of the elements in the field, that is, $f_x(R) =
x_{k-1}\,R^{k-1} + \cdots + x_1\,R + x_0$
and $f_y(R) =
y_{k-1}\,R^{k-1} + \cdots + y_1\,R + y_0$.
The first step is to
multiply the two polynomials $f_x$ and $f_y$. We can use different
polynomial multiplication algorithms depending on the value of $k$. 
Let us look at the expansion of $f_u$. Recall that 
taking a polynomial  modulo by $R^n + 1$ means replacing every 
occurrence of $R^n$ by $-1$.

\begin{eqnarray*}
	f_u(R) &=& f_x(R) \cdot f_y(R) \mod (R^k+1) \\
	&=& \sum_{m = 0}^{2\, k - 2} \sum_{0 \le i,j < k}^{i + j = m}x_i\,y_j\,R^m \mod (R^k + 1)\\
	&=& (x_{k-1}\,y_0 + x_{k-2}\,y_1 + x_{k - 3}\, y_2 + \cdots + x_1\,y_{k-2} + x_0\,y_{k-1})\,R^{k-1} \\
	&+& (x_{k-2}\,y_0 + x_{k-3}\,y_1 + \cdots + x_1\,y_{k-3} + x_0\,y_{k-2} - x_{k-1}\,y_{k-2})\,R^{k - 2}\\
	&+& (x_{k-3}\,y_0 + x_{k-4}\,y_1 + \cdots + x_0\,y_{k-3} - x_{k-1}\,y_{k-2} - x_{k-2}\,y_{k-1})\,R^{k - 2}\\	
	&& \ldots \\
	&+& (x_1\,y_0 + x_0\,y_1 - x_{k - 1}\,y_2 - \cdots - x_2\,y_{k - 1})\,R\\
	&+& (x_0\,y_0 - x_{k - 1}\,y_1 - \cdots - x_1\,y_{k - 1})\\
	&=& \sum_{m = 0}^{k - 1} (\sum_{0 \le i,j < k}^{i + j = m}x_i\,y_j - \sum_{0 \le i,j < k}^{i + j = k + m}x_i\,y_j)\,R^m\\
\end{eqnarray*}

Each coefficient $u_i$ of $f_u$ is the combination of $k$ monomials, so the
absolute value of each $u_i$ is bounded over 
by $k \cdot r^2$ which implies that it needs at most 
$\lfloor \log{k} + 2\,\log{r} \rfloor + 1$ bits to be encoded.
Since $k$ is usually between 4 to
256, a radix $r$ representation of $u_i$ of length 3 is 
sufficient to encode $u_i$.
Hence, we denote by $[c_i,h_i,l_i]$ the 3 integers uniquely given by:
\begin{enumerate}
\item $u_i = c_ir^2+h_ir+l_i$,
\item $0 \le h_i, l_i < r$.
\item $c_i \in [-(k-1),k]$,
\item $c_i u_i \geq 0$ holds.
\end{enumerate}
Then, we can rewrite:
\begin{eqnarray*}
	f_u(R) &=& f_x(R) \cdot f_y(R) \mod (R^k+1) \\
	&=&(c_0R^2+h_0R+l_0)+(c_1R^2+h_1R+l_1)R+(c_2R^2+h_2R+l_2)R^2+\cdots \\
	&& + (c_{k-2}R^2+h_{k-2}R+l_{k-2})R^{k-2}+(c_{k-1}R^2+h_{k-1}R+l_{k-1})R^{k-1}\\
	&=& \sum_{i = 0}^{k-1} (c_i\,R^{2+i} + h_i\,R^{1 + i} + l_i\,R^i)
\end{eqnarray*}

Now we obtain three vectors $\vec{c} = [c_0, c_1, \ldots, c_{k-1}]$,
$\vec{h} = [h_0, h_1, \ldots, h_{k-1}]$ and $\vec{l} = [l_0, l_1,
  \ldots, l_{k-1}]$ with $k$ coefficients each. As we shift $\vec{c}$ to
the right twice and $\vec{h}$ to the right once, we deduce three
numbers $c, h ,l$ in the radix-$r$ representation.

\begin{eqnarray*}
	c &=& c_{k-3}\,r^{k-1} + c_{k-4}\,r^{k-2} + \cdots + c_0\,r^2 + c_{k-2}\,r + c_{k-1}\\
	h &=& h_{k-2}\,r^{k-1} + h_{k-3}\,r^{k-2} + \cdots + h_1\,r^2 + h_0\,r + h_{k-1}\\
	l &=& l_{k-1}\,r^{k-1} + l_{k-2}\,r^{k-2} + \cdots + l_2,r^2 + l_1\,r + l_0\\
\end{eqnarray*}

At last we need two additions in $\mathbb{Z}/p\mathbb{Z}$ to compute the
result $u = c + h + l = x\,y \mod p$ with $x,y,u \in
\mathbb{Z}/p\mathbb{Z}$.

Now we consider the question of how to calculate $[l,h,c]$ quickly.
Because of the special structure
of $r$, where only two bits are 1, we can use some shift operations to
reduce the bit complexity and save on the cost of divisions.
Different $r$'s have different non-zero bits, 
but for clarity of presentation we use a particular radix $r$,
namely $r = 2^{63} + 2^{34}$, for the prime $P = r^k + 1$ with $k=8$.

Let $x_i, y_j$ be any two digits in the radix $r$ representation
of $x, y \in \mathbb{Z}/p\mathbb{Z}$.
Since $0 \le x_i,y_j \le r$ holds, we have
\begin{eqnarray*}
x_i\,y_j &=& (x_{i0} + x_i1\,r)\,(y_{j0} + y_j1\,r)\\
&=& x_{i0}\,y_{j0} + (x_{i0}\,y_{j1} + x_{i1}\,y_{j0})\,r + x_{i1}\,y_{j1}\,r^2
\end{eqnarray*}
where $0 \le x_{i0},y_{j0} < r$, and $x_{i1},y_{j1} \in \{0,1\}$.  Hence, we have $0 \le x_{i0}y_{j1},\, x_{i1}y_{j0}, \,x_{i1}\,y_{j1} < r$. We only need to consider the case of $x_{i0}\,y_{j0}$, where $0 \le x_{i0}\,y_{j0} <r^2 < 2^{127}$.
We can rewrite
\begin{eqnarray*}
	x_{i0} y_{j0} &=& (a_0+a_12^{32})(b_0+b_12^{32})\\
	&=&a_0b_0+a_0b_12^{32}+a_1b_02^{32}+a_1b_12^{64}\\
	&=&c_0+c_12^{64}.
\end{eqnarray*}

Notice that $a_0,a_1,b_0,b_1$ are in $[0,2^{32})$, using addition and shift operation, we can rewrite $x_{i0} y_{j0}$ into the form $c_0+c_12^{64}$, where $c_0 < 2^{64}$ and $c_1 < 2^{63}$. 
Then, we have:
\begin{eqnarray*}
	x_{i0} y_{j0}&=&c_0+c_12^{64}\\
	&=&c_0 + c_1^{\prime}2^{63} \quad\quad\quad\quad\quad\quad\quad\quad {\rm where}\,\,\,c_1^{\prime}=2 \, c_1, 0 \le c_1^{\prime} < 2^{64}\\
	&=&c_0 + c_1^{\prime}(2^{63}+2^{34}) - c_1^{\prime}2^{34}\\
	&=&c_0 + c_1^{\prime}r - c_1^{\prime}2^{34},
\end{eqnarray*}
where the part $c_0 + c_1^{\prime}r$ can be rewritten into the form of $l + hr + cr^2$ easily.

For $c_1^{\prime}2^{34}$, where $0\le c_1^{\prime} < 2^{64}$ holds, we observe:
\begin{eqnarray*}
	c_1^{\prime}2^{34} &=& (d_0+d_12^{29})2^{34} \quad\quad\quad\quad\quad\quad\quad\quad {\rm with} \,\,\,0\le d_0 < 2^{29}, 0\le d_1 < 2^{35}\\
	&=& d_02^{34} + d_12^{63}\\
	&=& d_02^{34} + d_1(2^{63}+2^{34}) - d_12^{34}\\
	&=& (d_0-d_1)2^{34} + d_1r\\
	&=& (e_0+e_12^{29})2^{34}+d_1r\quad\quad\quad\quad\quad\quad {\rm with} \,\,\,|e_0|<2^{29}, |e_1|<2^6\\
	&=& (e_0-e_1)2^{34}+e_1r+d_1r.
\end{eqnarray*}
Since $|(e_0-e_1)2^{34}| < r$ holds, 
the number $c_1^{\prime}2^{34}$ 
can easily be rewritten into the form of $l + hr + cr^2$. We add the $(l,h,c)$-representations of each part together, with some normalization we can get the result we need where $x_i\,y_j = l + h\,r + c\,r^2$. 

To summarize, the algorithm below 
uses only addition and shift operation 
to compute the $(l, h, c)$-representation of $x_i y_j$.
for $0 \le x_,y_j < 2^{64}$, and $0 \le l,h < r$, and $c \in \{0,1\}$.

\begin{algorithm}[H]
	\caption{An algorithm for rewriting $x_i y_j$ into $l+hr+cr^2$}
	\begin{algorithmic}[1]
		\Procedure{Rewrite}{$[l,h,c]=[x_i,y_j]$}
		\If{$x_i \ge r$}
		\State $x_{i1} := 1$
		\State $x_{i0} := x_i - r$
		\Else
		\State $x_{i1} := 0$
		\State $x_{i0} := x_i$
		\EndIf
		\Comment{$x_i =x_{i0} + x_{i1}r$}
		\If{$y_i \ge r$}
		\State $y_{i1} := 1$
		\State $y_{i0} := y_i - r$
		\Else
		\State $y_{i1} := 0$
		\State $y_{i0} := y_i$
		\EndIf
		\Comment{$y_i =y_{i0} + y_{i1}r$}
		\\
		\State $[v_1,v_2,v_3] := [0,x_{i0}\,y_{i1},0]$; \Comment{$x_{i0}y_{i1}r$}
		\State $[v_4,v_5,v_6] := [0,x_{i1}\,y_{i0},0]$; \Comment{$x_{i1}y_{i0}r$}
		\State $[v_7,v_8,v_9] := [0,0,x_{i1}\,y_{i1}]$; \Comment{$x_{i1}y_{i1}r^2$}
		\\
		\State $c_0 := x_{i0}\,y_{i1} - 2^{64}$
		\State $c_1 := (x_{i0}\,y_{i1}) >> 64$
		\State $c_1' := 2\,c_1$
		\Comment{$x_{i0}y_{i0}=c_0+c_12^{64} = c_0 + c_1'\,2^{63}$}
		\If{$c_0 \ge r$}
		\State $[v_{10},v_{11},v_{12}] := [c_0 - r,1,0]$
		\Else
		\State $[v_{10},v_{11},v_{12}] := [c_0,0,0]$
		\EndIf
		\Comment{$c_0=v_{10}+v_{11}r+v_{12}r^2$}
		\If{$c_1' \ge r$}
		\State $[v_{13},v_{14},v_{15}] := [0,c_1'-r,1]$
		\Else
		\State $[v_{13},v_{14},v_{15}] := [0,c_1',0]$
		\EndIf
		\Comment{$c_1'\,r=v_{13}+v_{14}r+v_{15}r^2$};
		\\
		\State $d_1 := c_1^{\prime}>>29$;
		\State $d_0 := c_1^{\prime}-d_1<<29$;
		\State $e_1 := (d_0-d_1)>>29$;
		\State $e_0 := (d_0-d_1-e_1<<29)$;
		\State $[v_{16},v_{17},v_{18}] := [(e_0-e_1)<<34,e_1+d_1,0]$;
		\\
		\State $[l,h,c] := [v_1+v_4+\cdots+v_{16},v_2+v_5+\cdots+v_{17},v_3+v_6+\cdots+v_{18}]$;
		\State \textbf{return} $[l,h,c]$;        
		\EndProcedure
	\end{algorithmic}
\end{algorithm}

The following algorithm calculates $u = x \, y \mod p$.
\begin{algorithm}[H]
	\caption{Computing 
		$u = x\,y \in {\mathbb{Z}}/p{\mathbb{Z}}$ for 
		$x, y \in {\mathbb{Z}}/p{\mathbb{Z}}$ using polynomial multiplication}
	\label{algo:polynomialmultiplication}
	\begin{algorithmic}[1]
		\Procedure{PolynomialMultiplication}{$\vec{x}, \vec{y}, r, k$}
		\State Multiply $f_u(R) = f_x(R) \cdot f_y(R) \mod (R^k + 1)$
		\For {m from 0 to k - 1}
		\State $[l_i,h_i,c_i]=\sum_{0 \le i,j < k}^{i + j = m}x_i\,y_j - \sum_{0 \le i,j < k}^{i + j = k + m}x_i\,y_j$
		\EndFor
		\\
		\Comment{The above k clauses can be executed in parallel}
		\\        
		\State $ShiftToRight[c_0, c_1,\ldots, c_{k-1}]$
		\State $ShiftToRight[c_{k-1}, c_0,\ldots, c_{k-2}]$
		\State $ShiftToRight[h_0, h_1,\ldots, h_{k-1}]$
		\State $u = c + h + l \mod p$
		\State \textbf{return} $u$
		\EndProcedure
	\end{algorithmic}
\end{algorithm}

\subsubsection{Modular multiplication based on integer multiplication}
\label{sec:integermultiplication}

This approach is more straight forward. For two numbers $x$ and $y$ in
our radix $r$ representation, we map the vectors $\vec{x}$ and
$\vec{y}$ to two polynomials $f_x, f_y \in \mathbb{Z}[R]$. Then we
evaluate the two polynomials at $r$, which gives us two integers $X$
and $Y$, using integer multiplication and modulo operation gives the
result $U = X\,Y \mod p$. At last, we only need to convert the product
back to the radix $r$ representation. See
Algorithm~\ref{algo:integermultiplication}.

\begin{algorithm}[h]
	\caption{Computing 
		$x\,y \in {\mathbb{Z}}/p{\mathbb{Z}}$ for 
		$x, y \in {\mathbb{Z}}/p{\mathbb{Z}}$ using integer multiplication}
	\label{algo:integermultiplication}
	\begin{algorithmic}[1]
		\Procedure{IntegerMultiplication}{$\vec{x}, \vec{y}, r, k, p$}
		\State $X:= 0 \, Y := 0$ \Comment{$X$ and $Y$ are GMP integers}
		\For{$i$ from $k - 1$ to $0$} 
		\State $X := X\cdot r + x_i$
		\State $Y := Y\cdot r + y_i$
		\EndFor
		\State $U := (X \cdot Y) \mod p$
		\State \textbf{return} GeneralizedFermatPrimeField(U)
		\EndProcedure
	\end{algorithmic}
\end{algorithm}

\subsection{Analysis}
\label{sec:analysisofmultiplication}

Here we want to analyze the complexity of multiplication in
$\mathbb{Z}/p\mathbb{Z}$, for $p = r^k + 1$, with radix $r$
representation. Since any number in our representation multiplied by
any power of $r$ is just a cyclic shift, we now only consider the case
that multiplication is between two arbitrary numbers, where both of
them are not powers of $r$.

In the following analysis, we compute $u = x \cdot y$, where $x =
x_{k-1} r^{k-1} + \cdots + x_0$ and $y = y_{k-1} r^{k-1} +\cdots +
y_0$ are two numbers in our Generalized Fermat Prime Field, with radix
$r$ representation. Let $\mathsf{M}$ be a multiplication time and let
$\omega$ be the number of bits in a machine word. We want to analyze
the complexity of multiplication with different approaches.

\subsubsection{Modular multiplication based on polynomial multiplication}
\label{sec:analysisofpolymul}
We view $x$ and $y$ as polynomials $f_x$ and $f_y$ in a variable $R$
with integer coefficients $x_0, \ldots, x_{k-1}$ and $y_0, \ldots
y_{k-1}$, whose bit sizes are at most that of one machine word. First
step in our multiplication is to multiply $f_x$ and $f_y$ in
$\mathbb{Z}[R]$, obtaining $f_u = u_{2k -2} R^{2k -2} + \cdots +
u_0$. The multiplication time of multiplying two polynomials of degree
less than  $k$
is $\mathsf{M}(k)$. The complexity of multiplying each pair of
coefficients is $\mathsf{M}(\omega)$ and the largest bit size of the
coefficients of $f_u$ is $\omega + k$, so the maximum complexity of
each operation in the polynomial multiplication is
${\max}(\mathsf{M}(\omega), \Theta(\omega + k))$, which gives us the
total complexity of this step:

\begin{eqnarray}
\mathsf{M}(k) \, {\max}(\mathsf{M}(\omega), \Theta(\omega + k)) \label{pm1}
\end{eqnarray}

In the next step, we compute the remainder of $f_u$ w.r.t $R^k + 1$. We should notice
that computing the remainder here is the same as computing  $f_u \mod (R^k + 1)$ that is
using $-1$ to replace every $R^k$. So, for each term in $f_u$, if the degree is greater than
$k-1$, reduce the degree by $k$ and reverse the sign for the coefficient. Combining the terms
with the same degree gives the final result of this step, $f_u = f_xf_y \mod (R^k + 1) = u_{k-1}R^{k-1} + \cdots + u_0$.
The total number of operations that we need to compute the remainder is in the
order of $\Theta(k)$, the bit complexity of each operation is $\Theta(\omega + k)$, thus the complexity of this step is:

\begin{eqnarray}
\Theta(k\,\omega) \label{pm2}
\end{eqnarray}

Next, we want to write each $u_i$ as $l_i + h_i r + c_i r^2$ with $0 \leq l_i,h_i,c_i < r$ using two divisions (one by $r^2$ and one by $r$), we get three vectors $[l_0, \ldots l_{k-1}]$, $[h_0, \ldots h_{k-1}]$ and $[c_0, \ldots c_{k-1}]$. Using cyclic shift on the three vectors, we obtain three numbers in radix r format: $z_l, z_h, z_c$. We need $2 \, k$ divisions in machine word size and three cyclic shifts for this step in total. So the complexity is:
\begin{eqnarray}
\Theta(k \, \mathsf{M}(\omega)) \label{pm3}
\end{eqnarray}

The last step in this approach is to add three numbers, $z_l, z_h, z_c$, together using two additions in $\mathbb{Z}/p\mathbb{Z}$. The complexity is:
\begin{eqnarray}
\Theta(k \, \omega) \label{pm4}
\end{eqnarray}

We can see that the second step has the greatest complexity \ref{pm2}. Thus, the total complexity of the approach based on polynomial multiplication is in the order of:
\begin{eqnarray}
\Theta(\mathsf{M}(k)) \, {\max}(\mathsf{M}(\omega), \Theta(\omega + k)) \label{pm}
\end{eqnarray}

\subsubsection{Modular multiplication based on reduction to integer multiplication}

In this approach, we convert two numbers in our radix $r$ representation $x$ and $y$ into two big integers $X$ and $Y$. Then we multiply them together as integers and convert the product to radix-$r$ representation. All of the operations we use in this method can be performed with the \href{https://gmplib.org/}{GMP} library~\cite{Granlund12}.

The GMP library chops the numbers into several parts which are called ``limbs". For numbers with different numbers of limbs, GMP uses different multiplication algorithms. Let us consider the case of multiplication between two equal size numbers with $N$ limbs each. For the base case with no threshold, the naive long multiplication is used with complexity of $\mathcal{O}(N^2)$. With the minimum of 10 limbs, GMP uses Karatsuba's algorithm with complexity of $\mathcal{O}(N^{\log3/\log2})$. Furthermore, multi-way Toom multiplication algorithms are introduced. Toom-3 is asymptotically $\mathcal{O}(N^{\log5/\log3})$, representing 5 recursive multiplies of $1/3$ original size each while Toom-4 has the complexity of $\mathcal{O}(N^{\log7/\log4})$. Though there seems an improvement over Karatsuba, Toom does more evaluation and interpolation so it will only show its advantage above a certain size. For higher degree Toom `n' half is used. Current GMP uses both Toom-6 `n' half and Toom-8 `n' half. At large to very large sizes, GMP uses a Fermat style FFT multiplication, following Sch{\"o}nhage and Strassen. Here $k$ is a parameter that controls the split, with FFT-k splitting the number into $2^k$ pieces, leading the complexity to $\mathcal{O}(N^{k/(k - 2)})$. It means $k = 7$ is the first FFT that is faster than Toom-3. Practically, the threshold for FFT in the GMP library is found in the range of $k = 8$, somewhere between 3000 and 10000 limbs(See more in \href{https://gmplib.org/}{GMP} library~\cite{Granlund12} manual).

Firstly, we reduce $x$ and $y$ to $X$ and $Y$ using the following method.

\begin{eqnarray}
X = (((x_{k-1}*r)+x_{k-2})*r \cdots +x_1)*r+x_0 
\end{eqnarray}

which needs $k-1$ additions and $k-1$ multiplications with at most $k\omega$ bits. Here, we still use $\mathsf{M}$ to represent the multiplication time. So, the complexity of this step is:
\begin{eqnarray}
\Theta(k\, \mathsf{M}(k\omega))
\end{eqnarray}

Then we multiply $X$ and $Y$ using operation from the GMP library. Let $U = X \cdot Y$. The complexity is 
\begin{eqnarray}
\mathsf{M}(k\omega)
\end{eqnarray}

At last, $U$ writes $u = u_{k -1} r^{k -1} + \cdots + u_0$ using $k-1$ divisions (by $r^{k-1}, \ldots ,r$). The complexity is:
\begin{eqnarray}
\Theta(k \, \mathsf{M}(k\omega)) 
\end{eqnarray}

The total complexity of this approach is
\begin{eqnarray}
\Theta(k \, \mathsf{M}(k\omega))
\end{eqnarray}

\subsection{Implementation with C code}
\label{sec:fft-based implementation}
In this section we give some details of how we actually implement the multiplication between two arbitrary elements in $\mathbb{Z}/p\mathbb{Z}$. We follow the basic idea of algorithm  \ref{algo:polynomialmultiplication} but there are more problems we need to solve.

Let $f_x(R),f_y(R)$ represent $x,y \in \mathbb{Z}/p\mathbb{Z}$ respectively. In the first step of the multiplication, we need to compute $f_u(R) = f_x(R)\cdot f_y(R) \mod (R^k + 1)$ in $\mathbb{Z}/p\mathbb{Z}$, which is a \textbf{Negacyclic convolution}. 
A convolution computes $f(x)\cdot g(x) \mod (x^n - 1)$ for two polynomials $f$ and $g$ with degree less than $n$. In \cite{MCA}, a fast algorithm~\ref{algo:convolution} of computing convolution is introduced.
\begin{algorithm}[H]
	\caption{Fast Convolution}
	\label{algo:convolution}
	\begin{algorithmic}[1]
		\State {\textbf{input:} \begin{itemize}
				\item[-] $n = 2^k \in \mathbb{N}$
				\item[-] two polynomials $f,g \in {\mathbb{A}}[x]$ with degree less than $n$,
				\item[-] a $n$-th primitive root of unity $\omega \in {\mathbb{A}}$.
			\end{itemize}
		}
		\State {\textbf{output:} \begin{itemize}
				\item[-] $f * g \in {\mathbb{A}}[x]$
			\end{itemize}
		}
		\Procedure{FastConvolution}{$f,g,\omega,n$}
		\State compute the first $n$ powers of $\omega$
		\State $\alpha := DFT_\omega(f)$
		\State $\beta := DFT_\omega(g)$
		\State $\gamma := \alpha \, \beta$
		\Comment Component-wise multiplication
		\State \textbf{return} $(DFT_\omega)^{-1}(\gamma) := \frac{1}{n}DFT_{\omega^{-1}}(\gamma)$
		\EndProcedure
	\end{algorithmic}
\end{algorithm}

 A similar approach can be used for computing the negacyclic convolution. 

Let $q$ be a prime, $\omega$ be an $n$-th primitive root of unity in  $\mathbb{Z}/q\mathbb{Z}$, and $\theta$ be a $2n$-th primitive root of unity in $\mathbb{Z}/q\mathbb{Z}$. Also we have two polynomials $f(x)$ and  $g(x)$ with degree less than $n$, we use $\vec{a}$ and $\vec{b}$ to represent the coefficient vector of the $f$ and $g$. First, we need to compute two vectors
\begin{equation}
\vec{A} = (1,\theta,\ldots,\theta^{n-1})
\end{equation}
and
\begin{equation}
\vec{A'} = (1,\theta^{-1},\ldots,\theta^{1 - n})
\end{equation}

The negacyclic convolution of $f$ and $g$ can be compute as follow
\begin{equation}
\vec{A'}\cdot\mathsf{InverseDFT}(\mathsf{DFT}(\vec{A}\cdot\vec{a}) \cdot \mathsf{DFT}(\vec{A}\cdot\vec{b}))
\end{equation}

All the dot multiplication between vectors are point-wise multiplication. The InverseDFT and DFTs are all $n$-point. We use unrolled inline DFTs in the implementation. The details of the DFTs are given in Section~\ref{sec:tensorFFT}.  This equation gives the following algorithm.

Algorithm \ref{algo:NegacyclicConvolution} is to compute $f_x(R) \cdot f_y(R) \mod (R^k + 1)$ over a finite field $\mathbb{Z}/q\mathbb{Z}$ with $q$ being a machine word size prime and $f_x(R)$, $f_y(R)$ being two polynomials of degree $k - 1$. $\vec{x}$ and $\vec{y}$ are the coefficient lists of $f_x$ and $f_y$.  
\begin{algorithm}[H]
	\caption{Computing $f_x(R)\cdot f_y(R) \mod (R^k + 1)$ in $\mathbb{Z}/q\mathbb{Z}$ using Negacyclic Convolution}
	\label{algo:NegacyclicConvolution}
	\begin{algorithmic}[1]
		\State {\textbf{input:} \begin{itemize}
				\item[-] a prime number $\tt q$ and $\tt k$ is a power of 2 with $k|(q - 1)$,
				\item[-] two vectors $\vec{\tt x}$ and $\vec{\tt y}$ of k elements,contain the coefficients of polynomials $f_x(R)$ and $f_y(R)$.
			\end{itemize}
		}
		\State {\textbf{output:} \begin{itemize}
				\item[-] a vector $\vec{\tt u}$ that contains the coefficients of polynomial $f_u(R) = f_x(R) \cdot f_y(R) \mod (R^k + 1)$ \end{itemize}}
		\Procedure{NegacyclicConvolution}{$\vec{x},\, \vec{y},\,q,\,k$}
		\State $\omega$ := PrimitiveRootOfUnity($q$,\,k);
		\Comment $\omega$ is the kth primitive root of unity of $q$
		\State $\theta$ := PrimitiveRootOfUnity($q$,\,2\,k);
		\Comment $\theta$ is the 2kth primitive root of unity of $q$
		\For {$0 \le i \le k - 1$}
		\State $A_i := \theta^i \mod q$;
		\State $x_i := x_i \cdot A_i \mod q$;
		\State $y_i := y_i \cdot A_i \mod q$;
		\EndFor
		\\
		\State $\vec{x} := \mathsf{DFT}(\vec{x},\omega,q,k)$;
		\State $\vec{y} := \mathsf{DFT}(\vec{y},\omega,q,k)$;
		\\        
		\For {$0 \le i \le k - 1$}
		\State $u_i := x_i \cdot y_i \mod q$;
		\EndFor
		\State $\vec{u} := \mathsf{DFT}(\vec{u},\omega^{-1} \mod q,q,k)$
		\For {$0 \le i \le k - 1$}
		\State $A'_i := \theta^{-i} \mod q$;
		\State $u_i := \frac{1}{k}\,(u_i \cdot A'_i) \mod q$;
		\EndFor
		\State \textbf{return} $\vec{u}$
		\EndProcedure
	\end{algorithmic}
\end{algorithm}

Notice that for $f_x$ and $f_y$ in our Generalized Fermat Prime Field $\mathbb{Z}/p\mathbb{Z}$, each coefficient is at most 63 bits. When computing $f_u(R) = f_x(R)\cdot f_y(R) \mod (R^k + 1)$, the size of the coefficients of $f_u$ can be at most $\log k + (2\cdot 63) = 126 + \log k$, which is more than one machine word, so that we cannot do the computation using single-precision arithmetic. But, multi-precision arithmetic can be very expensive and would make the algorithm inefficient. So we use two machine word negacyclic convolution in stead of one using big numbers. Hence, we need to apply the Chinese Remainder Theorem (CRT) to get the result that we want.

Let $p_1$ and $p_2$ be two machine word size prime numbers, so that we have $\mathsf{GCD}(p_1,p_2) = 1$. Then we use the extended Euclidean division to get $m_1$ and $m_2$ that satisfy the following relation
\begin{equation}
p_1\,m_1 + p_2\,m_2 = 1
\end{equation} 

Let $a$ be an integer and we have
\begin{eqnarray}
a_1 \equiv a \mod p_1\\
a_2 \equiv a \mod p_2
\end{eqnarray}
Then we compute $a \mod (p_1\,p_2)$ by
\begin{eqnarray}
	a &\equiv& a_2\,p_1\,m_1 + a_1\,p_2\,m_2 \mod (p_1\,p_2)\\
	&=& ((a_2\,m_1)\mod p_2)\,p_1 + ((a_1\,m_2)\mod p_1)\,p_2 \label{eq:crt}	
\end{eqnarray}

Hence, for $x,y \in \mathbb{Z}/p\mathbb{Z}$, we compute $u_1 = x \cdot y \mod p_1$ and $u_2 = x \cdot y \mod p_2$, then use \ref{eq:crt} to compute $u = x\cdot y \mod (p_1\,p_2)$. With some normalization we will get $u = x\cdot y \in \mathbb{Z}$. Let $R = k\,r^2$ be the upper bound of $(|u_0|,\ldots,|u_{k-1}|) \in \mathbb{Z}$. To get the correct answer, we need the following restrictions:

\begin{itemize}
	\item[1.] $R \le \frac{p_1\,p_2-1}{2}$
	\item[2.] the results we get from the CRT should be normalized so that they fall into the range of $[-\frac{p_1p_2 - 1}{2},\frac{p_1p_2 - 1}{2}]$
\end{itemize}

If $\frac{p_1\,p_2-1}{2}< R$, any result that is in the range of $(\frac{p_1\,p_2-1}{2},R)$ and $(-R,-\frac{p_1\,p_2-1}{2})$ will be inaccurate since the modular operation will make it in the range of $[-\frac{p_1\,p_2-1}{2},\frac{p_1\,p_2-1}{2}]$. 

As we mentioned before, all the results are in the range of $(-R,R)$ in $\mathbb{Z}$, which means $-\frac{p_1p_2 - 1}{2} < u_i < \frac{p_1p_2 - 1}{2}$ hold. Hence, after all the normalization we will have all the results in $\mathbb{Z}$ without losing any accuracy.

The small primes $p_1$ and $p_2$ are hard coded into the algorithm for now, where both $p_1 = 4179340454199820289$ and $p_2 = 2485986994308513793$ are 61-bit numbers. So, when choosing the Generalized Fermat prime, we should be very careful because of the two restrictions. For these two primes $p_1$ and $p_2$, the size of the chosen Generalized Fermat prime number $p = r^k + 1$ should be as follows:
\begin{eqnarray}
\log \frac{p_1 p_2 - 1}{2} > \log (k\,r^2)\\
121 > \log k + 2\log r\\
\log r < 59\,\,\,when\,\,\,k = 8\\
\log r < 58\,\,\,when\,\,\,k = 16\\
\log r < 58\,\,\,when\,\,\,k = 32\\
\log r < 57\,\,\,when\,\,\,k = 64
\end{eqnarray}

\bigskip
As we know, the modular operation in \ref{eq:crt} is expensive, so in the implementation we use what is called \textbf{reciprocal division} to reduce the cost of the modular operations. 

Let's say we want to compute $a \mod n$, instead of doing one single modular operation, we pre-compute the value of $ninv = 1/n$. Then we compute the result by
\begin{equation}
 a - n\cdot a\cdot ninv \equiv a \mod n
\end{equation}
Here, we only keep the integer part of $a \cdot ninv$, so that $n \cdot a \cdot ninv$ gives the quotient of the Euclidean division of $a$ by $n$. 

The following C code give the function of an efficient modular operation using the reciprocal division method.

\begin{lstlisting}[caption={Modular function using reciprocal division},captionpos=b,label = {lst:u64modu64}] 
void u64_mod_u64(usfixn64 &a, const usfixn64 &n){
	//a = a % n;
	double ninv = 1 / (double) n;
	usfixn64 q = (usfixn64) ((((double) a)) * ninv);
	usfixn64 res;
	res = a - q * n;
	a = res & (U64_MASK);
}
\end{lstlisting}   

\bigskip
Unlike modular operation, multiplication between two machine word size number sometimes can cause overflow, but using multi-precision numbers such as the ones given in the GMP library \cite{Granlund12} decreases the efficiency. To avoid that, we use two 64-bit numbers to represent the result of multiplication since the size of the result will be at most 128 bits. Let's say the sizes of $a$ and $b$ are at most 64 bits, we compute the multiplication between $a$ and $b$ by 
\begin{equation}
	s = a \cdot b = s_1\cdot 2^{64} + s_0
\end{equation}
where both of $s_1$ and $s_0$ are less than $2^{64}$.

To make the process even more efficient, we use assembly language in the following function.

\begin{lstlisting}[caption={Multiplication between two 64-bit numbers},captionpos=b,label = {lst:u64mulu64}] 
void __inline__ mult_u64_u64(const usfixn64 & a, const usfixn64 & b,
usfixn64& s0, usfixn64 &s1){
	//	__int128 mult = (__int128) a * (__int128) b;
	//	s0 = mult & (U64_MASK);
	//	s1 = mult >> 64;

	__asm__ (
		"movq  %2, %%rax;\n\t"          // rax = a
		"mulq %3;\n\t"// rdx:rax = a * b
		"movq  %%rax, %0;\n\t"// s0 = rax
		"movq  %%rdx, %1;\n\t"// s1 = rdx
		: "=rm" (s0),"=rm"(s1)
		: "rm"(a), "rm"(b)
		: "%rax", "%rdx");
}
\end{lstlisting}

We use function \ref{lst:u64mulu64} to compute the  $[t_0,t_1] = a_1\,m_2$ in equation \ref{eq:crt}. Then we need to do the modular by $p_1$. We can use a similar method as function \ref{lst:u64modu64}, but all the numbers will be in the size of 128 bits, so we use the representation of $s_1\,2^{64} + s_0$. 

To keep $1/p_1$ in the correct precision, we multiply it by $2^{128}$, and then we get
\begin{eqnarray}
\frac{2^{128}}{p_1} = p_1\_q\,2^{64} + p_1\_m
\end{eqnarray}

We have a function $\mathsf{mult\_u128\_u128\_hi128}$(see Appendix \ref{AppA}, function \ref{lst:u128mulu128}) to multiply $[t_0,t_1]$ and $[p_1\_q,p_1\_m]$ keeping the higher 64 bits only, which give the quotient $q_0$ of $a_1\,m_2$ divided by $p_1$. Then we have
\begin{equation}
a_1\,m_2 \mod p_1 = a_1\,m_2 - q_0\,p_1
\end{equation}

Again we use function \ref{lst:u64mulu64} to get the result of $(a_1\,m_2 \mod p_1)\cdot p_2$. Then use the same process to compute $(a_2\,m_1 \mod p_2)\cdot p_1$. Adding the two parts together gives us the final result of equation \ref{eq:crt}.

Using the same notation as above, the following algorithm  computes equation \ref{eq:crt} without using any multi-precision number. The corresponding C code can be found in Appendix \ref{AppA}.

\begin{algorithm}[H]
	\caption{Chinese Remainder Algorithm computing equation \ref{eq:crt}}
	\label{algo:crt}
	\begin{algorithmic}[1]
		\State {\textbf{input:} \begin{itemize}
				\item[-] two machine word size prime numbers $\mathtt{p_1}$ and $\mathtt{p_2}$,
				\item[-] $\mathtt{m_1}$ and $\mathtt{m_2}$ such that $p_1\,m_1 + p_2\,m_2 = 1$ holds,
				\item[-] $\mathtt{a_1}$ and $\mathtt{a_2}$ such that $a_1 \equiv a \mod p_1$ and $a_2 \equiv a \mod p_2$ hold.
			\end{itemize}
		}
		\State {\textbf{output:} \begin{itemize}
				\item[-] $a \mod (p_1p_2) = s_1\,2^{64} + s_0$ represented by $\mathtt{[s_0,s_1]}$. 
			\end{itemize}}
		\Procedure{CRT}{$p_1,p_2,m_1,m_2,a_1,a_2$}
		\State $[p_1\_q,p_1\_m]: = \frac{2^{128}}{p_1}$
		\State $[p_2\_q,p_2\_m]: = \frac{2^{128}}{p_2}$
		\State $[t_0,t_1] := \mathsf{multi\_u64\_u64}(a_1,m_2)$
		\State $[t_2,t_3] := \mathsf{multi\_u64\_u64}(a_2,m_1)$
		\State $q_0 := \mathsf{mult\_u128\_u128\_hi128}(t_0,t_1,p_1\_q,p_1\_m)$
		\State $q_1 := \mathsf{mult\_u128\_u128\_hi128}(t_2,t_3,p_2\_q,p_2\_m)$
		\State $[b_0,b_1] := \mathsf{multi\_u64\_u64}( q_0,p_1)$
		\State $[b_2,b_3] := \mathsf{multi\_u64\_u64}( q_1,p_2)$
		\State $c_1 := [t_0,t_1] - [b_0,b_1]$
		\State $c_2 := [t_2,t_3] - [b_2,b_3]$
		\State $[s_0,s_1] := \mathsf{multi\_u64\_u64}(c_0,p_2) + \mathsf{multi\_u64\_u64}(c_1,p_1)$
		\State Normalization $[s_0,s_1] \in [-\frac{p_1p_2 - 1}{2},\frac{p_1p_2 - 1}{2}]$ 
		\State \textbf{return} $[s_0,s_1]$
		\EndProcedure
	\end{algorithmic}
\end{algorithm}

\bigskip
After the negacyclic convolutions and the Chinese Remainder algorithm, we have $f_u = f_x \cdot f_y \mod (R^k + 1) \in \mathbb{Z}$. Next, we need to convert the coefficients of $f_u$ into the $(l,h,c)$ representation as we discussed in Section \ref{sec:fft-based}. 

Let $u_i = s_1\,2^{64} + s_0$ and $r$ be the radix of our Generalized Fermat Prime Field, we use a function $\mathsf{div\_by\_const\_R}$ (see Appendix \ref{AppA} function \ref{lst:divbyr}) to get $[m_0,q_0],[m_1,q_1]$ and $[m_2,q_2]$ that satisfy the following relation
\begin{eqnarray}
s_0 = q_0\,r + m_0 \,\,\, with\,\,\,q_0,m_0 < r\\
s_1 = q_1\,r + m_1\,\,\, with\,\,\,q_1,m_1 < r\\
2^{64} = q_2\,r + m_2\,\,\, with\,\,\,q_2,m_2 < r
\end{eqnarray}

Then we compute the $[l,h,c]$ by
\begin{eqnarray}
[l,h,c] &=& (q_0\,r + m_0) + (q_1\,r + m_1)\,(q_2\,r + m_2)\\
&=& q_1\,q_2\,r^2 + (m_1\,q_2 + m_2\,q_1 + q_0)\,r + (m_0 + m_1\,m_2)\\
&=& c'\,r^2 + h'\,r + l'
\end{eqnarray}

Notice that the $[l',h',c']$ we get here is not the final result yet since $h' = m_1\,q_2 + m_2\,q_1 + q_0$ and $l' = m_0 + m_1\,m_2$can be greater than $r$. We call function $\mathsf{div\_by\_const\_R}$ on $h'$ and $l'$ to normalize the result and give us $[l',h',c'] = [l_1,h_1,c_1]r + [l_0,h_0,c_0]$. We use addition with carry to get the final result $[l,h,c] = [l_1,h_1,c_1] + [l_0,h_0,c_0]$.

The following algorithm takes two numbers $[s_0,s_1]$ less than 64 bits as input, and output the $[l,h,c]$ as we defined in Section \ref{sec:fft-based}. The corresponding C code can be found in Appendix \ref{AppA} function \ref{lst:lhc}.

\begin{algorithm}[H]
	\caption{Computing $s_1\,2^{64} + s_0 = l + h\,r + c\,r^2$}
	\label{algo:lhc}
	\begin{algorithmic}[1]
		\State {\textbf{input:} \begin{itemize}
				\item[-] two machine word size numbers $\mathtt{s_1}$ and $\mathtt{s_0}$,
				\item[-] the radix $\mathtt{r}$.
			\end{itemize}
		}
		\State {\textbf{output:} \begin{itemize}
				\item[-] $[l,h,c]$ such that $s_1\,2^{64} + s_0 = l + h\,r + c\,r^2$. 
		\end{itemize}}
		\Procedure{LHC}{$s_1,s_0,r$}
		\State $[q_0,m_0] := \mathsf{div\_by\_const\_R}(s_0,r)$
		\State $[q_1,m_1] := \mathsf{div\_by\_const\_R}(s_1,r)$
		\State $[q_2,m_2] := \mathsf{div\_by\_const\_R}(2^{64},r)$
		\State $[l',h',c'] := (q_0\,r + m_0) + (q_1\,r + m_1)\,(q_2\,r + m_2)$
		\State $[l_0,l_1] := \mathsf{div\_by\_const\_R}(l',r)$
		\State $[h_0,h_1] := \mathsf{div\_by\_const\_R}(h',r)$
		\State $[c_0,c_1] := \mathsf{div\_by\_const\_R}(c',r)$
		\State $[l,h,c] := [l_0,h_0,c_0] + [l_1,h_1,c_1]$
		\State \textbf{return} $[l,h,c]$
		\EndProcedure
	\end{algorithmic}
\end{algorithm}

\bigskip
Now, we have all the coefficients of $f_u$ in the form of $[l,h,c]$. Rearranging the $k$ $[l,h,c]$ vectors gives us three vectors $\vec{l} = [l_0,\dots,l_{k-1}],\vec{h} = [h_0,\dots,h_{k-1}]$ and $\vec{c} = [c_0,\dots,c_{k-1}]$. Then we use function \ref{lst:MulPowRGFPF} to multiply $\vec{c}$ by $r^2$ and $\vec{h}$ by $r$. Finally, we use function \ref{lst:AdditionInGeneralizedFermatprimefield} to add $\vec{l},\vec{h},\vec{c}$  together to get the final result of $x\,y \in \mathbb{Z}/p\mathbb{Z}$. 

We call the this approach of multiplying two arbitrary elements in $\mathbb{Z}/p\mathbb{Z}$ the FFT-based multiplication in the Generalized Fermat Prime Field (FFT-based multiplication). The complete algorithm is as follow.

\begin{algorithm}[H]
	\caption{FFT-based multiplication for two arbitrary elements in $\mathbb{Z}/p\mathbb{Z}$}
	\label{algo:fft-based}
	\begin{algorithmic}[1]
		\State {\textbf{input:} \begin{itemize}
				\item[-] two vectors $\mathtt{\vec{x}}$ and $\mathtt{\vec{y}}$ representing the two elements $x$ and $y$ in $\mathbb{Z}/p\mathbb{Z}$,
				\item[-] two number $r$ and $k$ such that $p = r^k + 1$ is a generalized Fermat number.
			\end{itemize}
		}
		\State {\textbf{output:} \begin{itemize}
				\item[-] a vector $\mathtt{\vec{u}}$ representing the result of $x\cdot y \in \mathbb{Z}/p\mathbb{Z}$. 
		\end{itemize}}
		\State {\textbf{constant value:} \begin{itemize}
				\item[-] two machine word size primes $p_1$ and $p_2$,
				\item[-] two numbers $m_1$ and $m_2$ such that $p_1\,m_1 + p_2\,m_2 = 1$ holds. 
		\end{itemize}}
		\Procedure{FFT-basedMultiplication}{$\vec{x},\vec{y},r,k$}
		\State $\vec{z_1} := \mathsf{NegacyclicConvolution}(\vec{x},\vec{y},p_1,k)$
		\State $\vec{z_2} := \mathsf{NegacyclicConvolution}(\vec{x},\vec{y},p_2,k)$
		\For{$0 \le i < k$}
		\State $[{s_0}_i,{s_1}_i] := \mathsf{CRT}(p_1,p_2,m_1,m_2,{z_1}_i,{z_2}_i)$
		\EndFor
		\For{$0 \le i < k$}
		\State $[l_i,h_i,c_i] := \mathsf{LHC}({s_0}_i,{s_1}_i,r)$
		\EndFor
		\State $\vec{c} := \mathsf{MulPowR}(\vec{c},2,k,r)$
		\State $\vec{h} := \mathsf{MulPowR}(\vec{h},1,k,r)$
		\State $\vec{u} := \mathsf{BigPrimeFieldAddition}(\vec{l},\vec{h},k,r)$
		\State $\vec{u} := \mathsf{BigPrimeFieldAddition}(\vec{u},\vec{c},k,r)$
		\State \textbf{return} $\vec{u}$
		\EndProcedure
	\end{algorithmic}
\end{algorithm}

There are a lot of single-precision modular multiplication in Algorithm \ref{algo:fft-based}, these modular arithmetic can be very expensive and decrease the efficiency of the whole algorithm, so we decide to use Montgomery multiplication \cite{montgomery} inside this process.

Montgomery multiplication is an algorithm for performing modular
multiplication. It was presented by Peter L. Montgomery in 1985
\cite{montgomery}. This algorithm can speed up modular
multiplication by avoiding division by the modulus without affecting
modular addition and subtraction.

For a modulo $p$, let $R$ be a number greater than $p$ that is coprime
to $p$. Assume also that $R$ is some power of 2; hence multiplication
and division by $R$ can be done by shifting (on a computer
using binary expansions for numbers); thus, they can be seen
as inexpensive operations to perform.
Since ${\gcd}(R,p) = 1$ holds, there exists a unique pair $(R', p')$ of 
integers satisfying the following relation:
\begin{equation}
RR' - pp' = 1
\end{equation}
with $0 < R' < p$ and $0 < p' < R$. So that we have $p' = -p^{-1} \mod R$.

For a non-negative integer $a$, where $0 \le a < Rp$,  
{\em Montgomery reduction} computes $c := aR^{-1} \mod p$ 
without division modulo $p$.
Indeed, we have:
\begin{equation}
\begin{array}{rclc}
m &=& ap' \mod R \, & {\rm for}  \ \ 0 \le m < R \\
c &=& (a +mp)/R     & 
\end{array}
\end{equation}
if $c \ge p$ holds, then $c := c - p$  is performed. 

As we can see Montgomery multiplication requires a special representation of the elements that is for an element $a \in \mathbb{Z}/q\mathbb{Z}$ where $q$ is a machine word size prime, we rewrite $a$ into $(aR \mod q)$ where $R$ is the next power of 2 that is larger than $q$. In this form, multiplication can be performed efficiently without effect addition and subtraction. The Montgomery multiplication algorithm we use is as follow, supposing the machine word size is 64 bits. 

\begin{algorithm}[H]
	\caption{Montgomery Multiplication in $\mathbb{Z}/q\mathbb{Z}$}
	\label{algo:montmul}
	\begin{algorithmic}[1]
		\State {\textbf{input:} \begin{itemize}
				\item[-] two numbers $a$ and $b$ in $\mathbb{Z}/q\mathbb{Z}$,
				\item[-] the machine word size prime $q$,
				\item[-] a number $q' = -q^{-1} \mod 2^{64}$
			\end{itemize}
		}
		\State {\textbf{output:} \begin{itemize}
				\item[-] a vector $\mathtt{\vec{u}}$ representing the result of $x\cdot y \in \mathbb{Z}/p\mathbb{Z}$. 
		\end{itemize}}
		\State {\textbf{constant value:} \begin{itemize}
				\item[-] $c = a\,b\,R^{-1} \mod q$ 
		\end{itemize}}
		\Procedure{MontgomeryMultiplication}{$a,b,q,q'$}
		\State $R := 2^{64} - 1$
		\State $c := a\,b$
		\State $d := c\,q'$
		\State $c := c + q\,(d \& R)$
		\Comment $\&$ is the bit-wise and operation
		\State $c := c >> 64$
		\Comment $>> x$ is shift $x$ bits to the right 
		\If{$c \ge q$}
		\State $c := c - q$
		\EndIf
		\State \textbf{return} $c$
		\EndProcedure
	\end{algorithmic}
\end{algorithm}
The C code of the Montgomery multiplication for 64-bit numbers in the BPAS library can be found in Appendix \ref{AppA} function \ref{lst:montmul} (by Svyatoslav Covanov).

Once we have the Montgomery multiplication function, the ``convert-in'' and ``convert-out'' process can be very simple. Let $a$ be an element in $\mathbb{Z}/q\mathbb{Z}$, converting $a$ to the Montgomery representation can be done using the following equation
\begin{equation}
a\,R \equiv \frac{a \cdot R^2}{R} \mod q = \mathsf{MontgomeryMultiplication}(a,R^2,q,q')
\end{equation}
and the converting out from the Montgomery representation can be done by
\begin{equation}
a \equiv \frac{a\,R \cdot 1}{R} \mod q = \mathsf{MontgomeryMultiplication}(a\,R,1,q,q')
\end{equation}

So far, we have the full implementation of FFT-based multiplication between two arbitrary elements in the Generalized Fermat Prime Field. As we mentioned before, we also have an implementation based on integer multiplication using the GMP library\cite{Granlund12} following Algorithm \ref{algo:integermultiplication}. The experiment results comparing the two implementations can be found in Section \ref{sec:Experimentation}.

\section{A generic implementation of FFT over prime fields}
\label{sec:FFT}
In Section \ref{sec:tensorFFT}, we first review the tensor algebra
formulation of FFT, following the presentation of~\cite{FranchettiP11}. 
In the same section, we also
recall how one can transform the recursive formulation of the six-step
DFT to an iterative version, where all DFTs are then performed on a
fixed {\em base-case} size.  In the context of Generalized Fermat
prime fields, this reduction allows to take advantage of the ``cheap''
multiplication introduced in Section~\ref{sec:multiplybypowerofr}.
Section \ref{sec:finitefields}
introduces the different finite fields that are implemented in the
{\em Basic Polynomial Algebra Subprograms}, also known as the 
BPAS library~\cite{BPAS}. For efficiency reasons and convenience purposes, fields with
the same functionalities are implemented in both C and C++
languages. In Section~\ref{sec:bpasfft}, we explain how we
implemented the FFT in the BPAS library following the method in
Section~\ref{sec:tensorFFT}. We show the template functions for
different steps in the FFT which can adapt to all the finite fields in
the BPAS library. Also, we will explain how we implement the DFT
base-cases for 8, 16, 32 and 64 points.

\subsection{The tensor algebra formulation of FFT}
\label{sec:tensorFFT}
In the section we review the tensor formulation of FFT. First we
define the tensor product of two matrices over a
field\cite{pan2011algorithmic}. 

\begin{Definition}
	Let $n,m,q,s$ be positive integers and let $A, B$ be two matrices over $\mathtt{h}$ 
	with respective formats $m\times n$ and $q\times s$. 
	The tensor (or Kronecker) product of $A$ by $B$ is an $mq\times ns$ matrix is denoted by $A\otimes B$ and defined by
	\begin{equation}
	\label{eq:TensorProduct}
	A \otimes B = 
	\begin{bmatrix}
	a_{11}B &\cdots& a_{1n}B\\
	\vdots & \ddots & \vdots\\
	a_{m1}B & \cdots & a_{mn}B
	\end{bmatrix}
	\end{equation}
\end{Definition}

For example, we have two matrices
\begin{align*}
A = 
\begin{bmatrix}
0 &  1\\
2 & 3 
\end{bmatrix}
B =
\begin{bmatrix}
1 &  2 \\
3 & 4 
\end{bmatrix}
\end{align*}
Then we have
\begin{align*}
A \otimes B= 
\begin{bmatrix}
0\cdot B & 1 \cdot B\\
2 \cdot B & 3 \cdot B
\end{bmatrix}
=
\begin{bmatrix}
0 & 0 & 1 & 2\\
0 & 0 & 3 & 4\\
2 & 4 & 3 & 6\\
6 & 8 & 9 & 12 
\end{bmatrix}
\end{align*}

\begin{Definition}
	For matrices $A$ and $B$, operator $\oplus$ is defined as follow
	\begin{align*}
	A \oplus B = 
	\begin{bmatrix}
	A &  0\\
	0 & B 
	\end{bmatrix}
	\end{align*}
	For $n$ matrices $A_0 \ldots A_{n-1}$, the $\oplus$ sum of them is defined as 
	\begin{equation}
	\label{eq:DirectSum2}
	\bigoplus_{i=0}^{n-1}A_i = A_0 \oplus A_1 \oplus \cdots \oplus A_{n-1} 
	=   \left[\begin{array}{cccc}
	A_0 &      &         &         \\
	&  A_1 &         &         \\
	&      &  \ddots &         \\
	&      &         & A_{n-1}
	\end{array}\right].
	\end{equation}
\end{Definition}

In a ring $R$, an $n$-point $\mathsf{DFT}_n$ can be seen as a linear map of $R^n \mapsto R^n$. In the BPAS library, we use the six-step recursive FFT algorithm presented in \cite{FranchettiP11}. It can be represented by the following equation
\begin{equation}
\label{eq:sixstepfft}
\mathsf{DFT}_N = L_K^N\,(I_J \otimes \mathsf{DFT}_K) L_J^N\, D_{K,J}\,(I_K \otimes \mathsf{DFT}_J)\,L_K^N\,\,\,with\,\,\,N = J\,K
\end{equation}
which uses the divide-and-conquer idea of F{\"u}rer's algorithm. For the part of $I_K \otimes \mathsf{DFT}_J$, we can further expand it to using the base-case $\mathsf{DFT}_K$. Hence, if we have an efficient implementation of the base-case, we will have an efficient algorithm for FFT.

In equation \ref{eq:sixstepfft}, $L_K^N$ is called a stride permutation and $D_{K,J}$ is called a twiddle factor. They are defined as follow.

\begin{Definition}
	\label{def::permutation}
	The stride permutation $L_{m}^{mn}$ permutes an input vector 
	$\vec{x}$ of length $mn$ as follows
	\begin{equation}
	\vec{x}[in + j] \mapsto \vec{x}[jm+i]
	\end{equation}
\end{Definition}

Basically what the stride permutation does is, for an input vector $\vec{x}$ with length $mn$, it treats the vector as a $n \times m$ matrix and does a transposition on it. 
\begin{equation}
L_m^{mn}(M_{n \times m}) = (M_{n \times m})^T
\end{equation}

For example, the input vector is $\vec{x}_8 = [0,1,2,3,4,5,6,7]$, with $m = 2$ and $n m = 8$, the $n \times m$ matrix is 
\begin{align*}
M_{n \times m}^T = 
\begin{bmatrix}
0 &  1\\
2 & 3 \\
4 & 5 \\
6 & 7
\end{bmatrix}^T
=
\begin{bmatrix}
0 &  2 & 4 & 6 \\
1 & 3 & 5 & 7
\end{bmatrix}
\end{align*}
So $L_m^{mn}(\vec{x}) = [0,2,4,6,1,3,5,7]$

\begin{Definition}
	\label{def:twiddle}
	The twiddle factor $D_{K,J}$ is a matrix of the powers of $\omega$. 
	\begin{equation}
	\label{eq:twiddle}
	D_{K,J} = 
	\bigoplus_{j = 0}^{K-1}\,diag\,(1,\omega_i^j,\ldots,\omega_i^{j(J-1)})
	\end{equation}
\end{Definition}

We can compute all the twiddle factor multiplication with Algorithm \ref{algo:fft-based}, but as is introduced in F{\"u}rer's paper\cite{DBLP:journals/siamcomp/Furer09}, we want to compute the base-case $\mathsf{DFT}_K$ using a cheaper multiplication with some $K$-th primitive roof of unity. 

Now, we want to compute $\mathsf{DFT}_{K^e}$ by computing $\mathsf{DFT}_K$. The twiddle factor here should be $D_{K,K^{e-s}}$ where $\omega_i = \omega^{K^{s-1}}$ for $(1 \le s < e)$. And we know from Section \ref{sec:GFPF} that for a Generalized Fermat prime $p = r^k + 1$, $r$ is a $2k$-th primitive root of unity, then we have $\omega^N = r^{2k} = 1 \mod p$. Hence, we can using following method to compute the twiddle factor multiplication $y = x \cdot \omega^{i(N/K) + j}$.
\begin{eqnarray}
y &=& (x \cdot \omega^{iN/K}) \cdot \omega^j\\
&=& (x \cdot r^{2ki/K})\cdot \omega^j\\
&=& (x \cdot r^i)\cdot \omega^j
\end{eqnarray}
We use Algorithm \ref{lst:MulPowRGFPF} to compute the multiplication with $r^i$ which is very cheap, and only compute the twiddle factor multiplication with $\omega^j$ using Algorithm \ref{algo:fft-based}. We can pre-compute all the power of $\omega^j$ for $0 \le j < N/K$ to further reduce the complexity of the algorithm. In conclusion, to compute $\mathsf{DFT}$ on $K^e$ points, we need to pre-compute the power of $\omega^j$ for all $0 \le j < K^{e-1}-1$.

We can see that once we have an efficient implementation of the
base-case $\mathsf{DFT}_K$, we can compute $\mathsf{DFT}_N$ at any
size where $N$ is some power of 2. In Section~\ref{sec:bpasfft}, we
will explain how we implement the efficient base-case in the BPAS
library.

\subsection{Finite fields in the BPAS library}
\label{sec:finitefields}

In order to provide both efficiency and convenience, we implemented the following finite fields in the BPAS library using either the C or C++ language.

\paragraph{{\tt SmallPrimeField} C++ Class:}
\label{smallprimefieldclass}
C++ implementation in the BPAS library of a
prime field of the form $\mathsf{GF}(p)$ 
where $p$ is an arbitrary prime number of machine word size.

\paragraph{{\tt SmallPrimeField} in C:}
\label{smallprimefieldc}
Set of C functions in the BPAS library implementing
arithmetic operations in a 
prime field of the form $\mathsf{GF}(p)$ where $p$ is an arbitrary prime number of machine word size.

\paragraph{{\tt BigPrimeField} C++ Class:}
\label{bigprimefieldclass}
C++ implementation in the BPAS library of a 
prime field of the form $\mathsf{GF}(p)$ where 
$p$ is an arbitrary prime number without any restrictions on its size.

\paragraph{{\tt BigPrimeField} in C:}
\label{bigprimefieldc}
Set of C functions  (provided by the GMP library) implementing
arithmetic operations in a
prime field of the form $\mathsf{GF}(p)$ 
where $p$ is an arbitrary prime number  without any restrictions on its size.

\paragraph{{\tt GeneralizedFermatPrimeField} C++ Class:}
\label{GFPFclass}
C++ implementation in the BPAS library of a 
prime field of the form $\mathsf{GF}(p)$ where $p$ is a Generalized
Fermat prime, see Section \ref{sec:GFPF}.

\paragraph{{\tt GeneralizedFermatPrimeField} in C (GMP-based):}
\label{GFPFgmpc}
Set of C functions implementing
arithmetic operations in a prime field of 
the form $\mathsf{GF}(p)$ where $p$ is a Generalized
Fermat prime, see Section  \ref{sec:GFPF}.

\paragraph{{\tt GeneralizedFermatPrimeField} in C (FFT-based):}
\label{GFPFfftc}
Set of C functions implementing
arithmetic operations in a prime field 
of the form $\mathsf{GF}(p)$ where $p$ is a Generalized
Fermat prime, see Section  \ref{sec:GFPF}. Note that in this case,
the multiplication of two elements of the field is done by FFT as we described in \ref{sec:fft-based implementation}.
\bigskip

Both of the {\tt SmallPrimeField} implementations use machine word size
primes (the {\tt long long int} type in C and C++) and have the same
functionalities. And all the arithmetic is done using Montgomery
representation, see ~\cite{montgomery}.
In the C++ class, we convert all the objects into
Montgomery representation in the constructor and convert out 
when users call the convert out method or printing method. The C
version has functions for converting in and out, the users should call
these functions before and after doing any computations.

Inside the {\tt SmallPrimeField} class, we overload the arithmetic
operators $+,-,*,/$ as well as the Boolean operators
$==,!=,>,<,>=,<=$; we also have methods for computing the inverse of
an elements in the finite field as well as for exponentiation by 
any integer exponent. 
For multiplication, we use Algorithm~\ref{algo:montmul}. Finally, we follow the
method introduced in~\cite{GuideToECC} (see Algorithms 2.23 and 2.25) for the
Montgomery-based inversion.

The calling sequence of the {\tt SmallPrimeField} 
class is as follows.
\begin{lstlisting}[caption={Calling sequence of {\tt SmallPrimeField} 
class in the BPAS library},captionpos=b,label = {lst:callspfclass}]
#include "bpas.h" 
int main(){
int p = 257;
SmallPrimeField::setPrime(p);
//set the prime to 257
int n = 234;
SmallPrimeField a(n);
//create an object that equal to n mod p
SmallPrimeField b(100);
//create an object that equal to 100 mod p
SmallPrimeField a;
//create a 0 object
c = a + b;
c = a - b;
c = a * b;
c = a.inverse();
c = a^5;
cout << c << endl;
}
\end{lstlisting}

An example of using the C implementation of {\tt SmallPrimeField} follows.
\begin{lstlisting}[caption={Calling sequence of {\tt SmallPrimeField} macro in the BPAS library},captionpos=b,label = {lst:callspfmacro}]
#include "bpas.h" 
int main(){
long int p = 257;
long int Pp = getPp(p,R);
//R can be computed as 2^64 mod p
long int a = 100;
long int b = 576;
a = covert_in(a, p,R);
b = covert_in(b, p,R);
a = add(a,b,p);
//a = a + b mod p
a = sub(a,b,p);
//a = a - b mod p
a = multi(a,b,p,R,Pp);
//a = a*b/R mod p;
a = covert_out(a, p,R);
}
\end{lstlisting}

Section \ref{sec:Experimentation} shows the experimental data of FFT over 
{\tt SmallPrimeField} in C and C++.

\bigskip
The {\tt BigPrimeField} class has the same functionality as the {\tt
	SmallPrimeField} class, except that all the arithmetic is done using
GMP integers (type mpz\_class). So users can choose prime numbers of 
any size.

\bigskip
The {\tt GeneralizedFermatPrimeField} Class and 
{\tt GeneralizedFermatPrimeField} C functions follow the representation and arithmetic we introduced in Section~\ref{sec:GFPF}. We implemented multiplication between two arbitrary element using both FFT-based method and GMP-based method in the C version. The default one for overloading the operator $*$ in the class is the GMP-based one.

\subsection{BPAS implementation of the FFT }
\label{sec:bpasfft}
In the BPAS library, we implemented an FFT algorithm using the six-step FFT we described in Section~\ref{sec:tensorFFT}. Recall the six-step FFT formula
\begin{equation*}
\mathsf{DFT}_N = L_K^N\,(I_J \otimes \mathsf{DFT}_K) L_J^N\, D_{K,J}\,(I_K \otimes \mathsf{DFT}_J)\,L_K^N\,\,\,with\,\,\,N = J\,K
\end{equation*}
where $L$ is the stride permutation, and $D$ is the twiddle factor multiplication.

Other than the three steps of the permutation and one call of twiddle factor
multiplication, we still need to perform the the base-case $\mathsf{DFT}_K$ as we
explained in Section~\ref{sec:tensorFFT}. Inside the BPAS library, we
implemented base-cases for $K=8,16,32,64$ and reduced them into
$\mathsf{DFT}_2$. First, let us see the function for computing
$\mathsf{DFT}_2(x_0,x_1)$
\begin{equation}
\mathsf{DFT}_2(x_0,x_1) = (x_0 + x_1,x_0 - x_1)
\end{equation}
For $K = 2^n$, we reduce $\mathsf{DFT}_K$ to $\mathsf{DFT}_2$ by
\begin{equation}
\mathsf{DFT}_{2^n} = L_2^{2^n}\,(I_{2^{n-1}} \otimes \mathsf{DFT}_2)\,L_{2^{n-1}}^{2^n}\,D_{2,2^{n-1}}\,(I_2 \otimes \mathsf{DFT}_{2^{n-1}})\,L_2^{2^n}
\end{equation}

We follow Algorithm~\ref{algo:dft_general} 
to compute a $N$-point DFTs where $N = K^e$ and $e$ is a positive
integer.

\begin{algorithm}[H]
	\caption{Computing DFT  on $K^e$ points in $\mathbb{Z}/p\mathbb{Z}$}
	\label{algo:dft_general}
	\begin{algorithmic}[1]
		\State {\textbf{input:} \begin{itemize}
				\item[-] size of the base-case $K$(8,16,32 or 64), a positive  integer $e$,
				\item[-] a vector $\vec{x}$ of size $K^e$,
				\item[-] $\omega$ which is a $K^e$-th primitive root of unity in $\mathbb{Z}/p\mathbb{Z}$.
			\end{itemize}
		}
		\State {\textbf{output:} \begin{itemize}
				\item[-] the final result stored in $\vec{x}$
			\end{itemize}
		}
		\Procedure{DFT\_general}{$\vec{x},K,e,\omega,$}
		\For{$0 \le i < e-1$}
		\For{$0 \le j < K^i$}
		\State stride\_permutation($\&x_{jK^{e-i}},K,K^{e-i-1}$)
		\EndFor
		\EndFor
		\Comment Step 1
		\State $\omega_a := \omega^{K^{e-1}}$
		\For{$0 \le j < K^{e-1}$}
		\State $idx:=jK$
		\State DFT\_K($\&x_{idx},\omega_a$)
		\EndFor
		\Comment Step 2
		\For{$e-2 \ge i \ge 0$}
		\State $\omega_i := \omega^{K^i}$
		\For{$0 \le j < K^i$}
		\State $idx := j\,K^{e-i}$
		\State twiddle($\&x_{idx},K^{e-i-1},K,\omega_i$)
		\Comment Step 3
		\State stride\_permutation($\&x_{idx},K^{e-i-1},K$)
		\Comment Step 4
		\EndFor
		\For{$0 \le j < K^{e-1}$}
		\State $idx := jK$
		\State DFT\_K($\&x_{idx},\omega_a$)
		\EndFor
		\Comment Step 5
		\For{$0 \le j < K^i$}
		\State $idx := jK^{e-i}$
		\State stride\_permutation($\&x_{idx},K,K^{e-i-1}$)
		\EndFor
		\EndFor
		\Comment Step 6
		\EndProcedure
	\end{algorithmic}
\end{algorithm}

The same code for stride permutation 
(function stride\_permutation in Algorithm \ref{algo:dft_general})
is used
for all BPAS finite fields. Indeed that part 
is independent of the finite field used for the FFT. 
The C code of the stride permutation is listed below.

\begin{lstlisting}[caption={Stride permutation for FFT},captionpos=b,label = {lst:permutation}]
void stride_permutation(ELEMENTS* A,int m, int n){
int blocksize=m^((m^n)&(-(m>n)));
blocksize=BLOCKSIZE^((BLOCKSIZE^blocksize)&(-(BLOCKSIZE>blocksize)));
ELEMENTS* B = new ELEMENTS[m*n];
for (int i = 0; i < n; i += blocksize) {
for (int j = 0; j < m; j += blocksize) {
// transpose the block beginning at [i,j]
for (int k = i; k < i + blocksize; ++k) {
for (int l = j; l < j + blocksize; ++l) {
B[k+l*n] = A[l+k*m];
}
}
}
}
for (long int i=0;i<m*n;i++)
A[i]=B[i];
}
\end{lstlisting}

The same template code for twiddle factor multiplication 
(function twiddle in Algorithm \ref{algo:dft_general}) is 
used for all BPAS finite fields. This template code
has 4  specializations
\begin{itemize}
	\item one for both {\tt SmallPrimeField} (C and C++); switching
	between C and C++ is done by compilation directive
	\item one for each of  {\tt BigPrimeField} (C and C++);  
	\item one for {\tt GeneralizedFermatPrimeField} (C and C++); switching
	between C and C++ is done by compilation directive.
\end{itemize}

The C code of the twiddle template function is as follows. 
The only difference for different prime fields is the multiplication used in line 5 and 6. 
\begin{lstlisting}[caption={Twiddle factor multiplication for FFT},captionpos=b,label = {lst:twiddle}]
void twiddle(ELEMENTS* vector, int m, int n, ELEMENTS omega_w){
for (int j=0;j<n;j++){
for(int i=0;i<m;i++){
ELEMENTS t;
t=POW(omega_w,(i*j));
vector[j*m+i]=vector[j*m+i]*(t);
}
}
}
\end{lstlisting}

For the base-case,
that is, DFT\_K in Algorithm~\ref{algo:dft_general},
the same template code for  is 
used for all BPAS finite fields. Similarly
to the function twiddle, specializations are provided
for each BPAS finite field. Three specializations
differ by their calls to functions doing
addition, subtraction and multiplication.
Note that  for multiplication by a power of the
primitive root, in the case of {\tt GeneralizedFermatPrimeField}, we
use the techniques described in Section~\ref{sec:multiplybypowerofr}

Now, let us consider the base-case of $K = 8$, where $\omega$ is an 8-th primitive root in $\mathsf{GF}(p)$. 
\begin{eqnarray}
\mathsf{DFT}_{8} &=& L_2^{8}\,(I_{4} \otimes \mathsf{DFT}_2)\,L_{4}^{8}\,D_{2,4}\,(I_2 \otimes \mathsf{DFT}_{4})\,L_2^{8}\\
\mathsf{DFT}_{4} &=& L_2^{4}\,(I_{2} \otimes \mathsf{DFT}_2)\,L_{2}^{4}\,D_{2,2}\,(I_2 \otimes \mathsf{DFT}_{2})\,L_2^{4}\\
\mathsf{DFT}_{8} &=& L_2^{8}\,(I_{4} \otimes \mathsf{DFT}_2)\,L_{4}^{8}\,D_{2,4}\,(I_2 \otimes ( L_2^{4}\,(I_{2} \otimes \mathsf{DFT}_2)\,L_{2}^{4}\,D_{2,2}\,(I_2 \otimes \mathsf{DFT}_{2})\,L_2^{4} )\,L_2^{8} \label{eq:dft8}
\end{eqnarray}
where
\begin{eqnarray}
D_{2,4} &=& (1,1,1,1,\omega_0^0,\omega_0^1,\omega_0^2,\omega_0^3)\\
D_{2,2} &=& (1,1,\omega_1^0,\omega_1^1)
\end{eqnarray}

For a prime field with an arbitrary $p$, we have for $\mathsf{DFT}_8$, $\omega_0  = \omega^{N/K} = \omega$ and for  $\mathsf{DFT}_4$, $\omega_1  = \omega^{(N/K)^2} = \omega^2$. 

For a Generalized Fermat prime field where the prime is $p = r^4 + 1$ 
we have for $\mathsf{DFT}_8$, $\omega_0  = \omega^{N/K} = r^{2k/K} = r$ and for  $\mathsf{DFT}_4$, $\omega_1  = \omega^{(N/K)^2} = r^2$. Then, the twiddle factors are
\begin{eqnarray}
D_{2,4} &=& (1,1,1,1,1,r,r^2,r^3)\\
D_{2,2} &=& (1,1,1,r^2)
\end{eqnarray}
Hence, multiplication with the twiddle factors can be done by cyclic shift from Section~\ref{sec:multiplybypowerofr}.

Now, we follow Equation (\ref{eq:dft8}) from right to left and get the following unrolled algorithm for $\mathsf{DFT}_8$.

\begin{algorithm}[H]
	\caption{Unrolled DFT base-case when $K = 8$}
	\label{algo:dft8}
	\begin{algorithmic}[1]
		
		\Procedure{DFT8}{$\vec{a}, \omega_i$}
		\State DFT2($\&a_0,\&a_4$); 
		\State DFT2($\&a_2,\&a_6$);
		\State DFT2($\&a_1,\&a_5$);
		\State DFT2($\&a_3,\&a_7$);
		\Comment  dft on permuted indexes
		\State 
		\State $a_6$ := $a_6\,\omega^2$;   
		\State $a_7$ := $a_7\,\omega^2$;
		\Comment twiddle
		\State 
		\State DFT2($\&a_0,\&a_2$); 
		\State DFT2($\&a_4,\&a_6$);
		\State DFT2($\&a_1,\&a_3$);
		\State DFT2($\&a_5,\&a_7$);
		\Comment dft on permuted indexes
		\State
		\State $a_5$ := $a_5\,\omega^1$;
		\State $a_3$ := $a_3\,\omega^2$;
		\State $a_7$ := $a_7\,\omega^2$;
		\Comment twiddle
		\State
		\State DFT2($\&a_0,\&a_1$);
		\State DFT2($\&a_4,\&a_5$);
		\State DFT2($\&a_2,\&a_3$);
		\State DFT2($\&a_6,\&a_7$);
		\Comment dft on permuted indexes
		\State
		\State swap($\&a_1,\&a_4$);
		\State swap($\&a_3,\&a_6$);
		\Comment final permutation	
		\State \textbf{return} $\vec{a}$;
		\EndProcedure
	\end{algorithmic}
\end{algorithm}

The $\mathsf{swap}$ function swap the value of of its two parameters. 
The other DFT base-case codes are relatively long so we only show the number of lines here. The numbers of lines for unrolled $\mathsf{DFT}_K$ are shown in Table~\ref{tab:linenumber} (not counting comments). The C code can be found in the BPAS library. 
\begin{table}[htbp]
	\centering%
	\renewcommand*{\arraystretch}{1.35}
	{
		\begin{scriptsize}
			\begin{tabular}{|l|l|l|l|l|} \hline 
				$K$ & 8 & 16 & 32 & 64 \\ \hline
				number of lines & 19 & 55 & 141 & 359 \\ \hline		
			\end{tabular}
		\end{scriptsize}
	}
	\caption{\small Numbers of lines in n-point unrolled FFT.}
	\label{tab:linenumber}
\end{table}

Finally, and consequently, the same template code for  Algorithm 5.1
is used for all BPAS finite fields.

\section{Experimentation}
\label{sec:Experimentation}
In this section, we present experimental data
of FFT over the finite fields in the BPAS library.
In Section~\ref{sec:fftoversmallprimefield}, we compare 
our implementation of FFT over {\tt SmallPrimeField} Class and 
C functions as well as another highly optimized FFT implementation 
from the BPAS library. 
Also, we compare the two implementations of the multiplication in $\mathbb{Z}/p\mathbb{Z}$ introduced in Section~\ref{GFPFMulti}; the results are in Section~\ref{sec:expofmultiplicationingfpf}.

In Section~\ref{sec:fftoverbigprimefield}, we report results of FFT over 
over {\tt BigPrimeField}, {\tt GeneralizedFermat\-PrimeField} 
using GMP-based multiplication and 
{\tt GeneralizedFermatPrimeField}  using 
FFT-based multiplication written in C.
Clearly, the latter scenario
gives better running tines than the other two.

All the experimental results have been verified using Maple, Python and GMP~\cite{Granlund12}.                        

\subsection{FFT over small prime fields}
\label{sec:fftoversmallprimefield}

Before the work reported in this article,
various implementations of FFT over small finite fields
were developed in the BPAS library.
In particular, a highly optimized version 
by Svyatoslav Covanov  is presented in~\cite{BPAS}.
For this latter, the source code of the FFT 
is generated at compile time: it takes
into account the characteristics of the targeted hardware 
and it is specialized for a particular
prime field. This latter feature allows 
compiler optimization strategies which
are not possible for a generic implementation
like the one presented in Section~\ref{sec:FFT}.

Nevertheless, it is interesting to compare our generic implementation
(over the {\tt SmallPrime\-Field} class and {\tt SmallPrimeField} in C)
against the highly optimized FFT produced by Covanov's code generator.

As introduced in Section~\ref{sec:bpasfft}, our implementation of FFT
is based on an unrolled code for the base-case DFT functions
$\mathsf{DFT}_K$, where $K$ can be 8,16,32 or 64. In the following
results, we refer to Svyatoslav Covanov's implementation as
\textit{Svyatoslav}, and refer to our implementation of FFT using
base-case $\mathsf{DFT}_K$ as $\mathsf{DFT}_K\_C++$ and
$\mathsf{DFT}_K\_C$ depending on which {\tt SmallPrimeField} (C++
class or C functions) it uses.

\begin{figure}[H]
	\centering
	\includegraphics[totalheight=6cm]{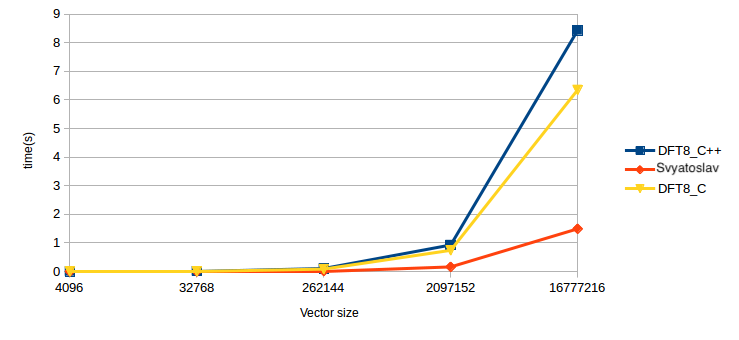}
	\caption{FFT over small prime field with $\mathsf{DFT}_8$}
	\label{fig:spf8}
\end{figure}

\begin{figure}[H]
	\centering
	\includegraphics[totalheight=6cm]{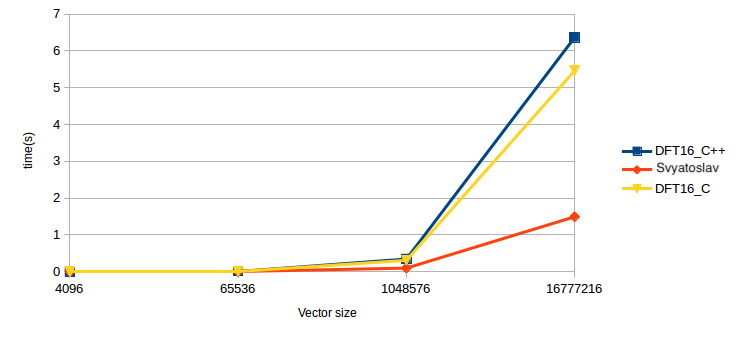}
	\caption{FFT over small prime field with $\mathsf{DFT}_{16}$}
	\label{fig:spf16}
\end{figure}

\begin{figure}[H]
	\centering
	\includegraphics[totalheight=6cm]{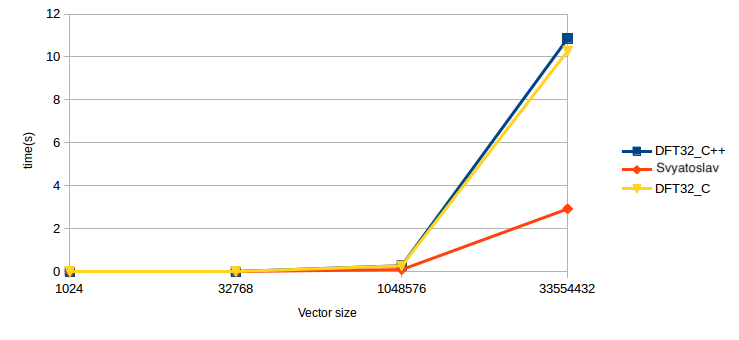}
	\caption{FFT over small prime field with $\mathsf{DFT}_{32}$}
	\label{fig:spf32}
\end{figure}

Figures~\ref{fig:spf8}, \ref{fig:spf16} and \ref{fig:spf32} show the time 
spending on FFT over large vectors using base-case size of 8, 16 and 32 
respectively. The $x$-axis gives the size of the vectors. 
The $y$-axis is time in seconds. All the results are based on average 
time of 50 trails. We can see that the C++ class is slower than the C functions with the implementation of the same algorithm.

Our best result is still slower than Svyatoslav's by the factor of 5. 
As mentioned above, this is because his code is specialized 
at the prime number together with embedded assembly code.
All the experimental results in this section were realized on an
Intel(R) Core(TM) 2.90GHz i5-528U CPU.


\subsection{Multiplication in generalized Fermat prime fields}
\label{sec:expofmultiplicationingfpf}

As in Section \ref{GFPFMulti}, we have two multiplication algorithms
between two arbitrary elements of the
generalized Fermat prime field
$\mathbb{Z}/p\mathbb{Z}$. One of
them is based on negacyclic convolution using unrolled DFT base-case
\ref{sec:fft-based} (referred to as FFT-based in the figures and
tables), the other one is based on GMP integer multiplication
\ref{sec:integermultiplication} (referred to as GMP-based in the
figures and tables). We want to compare the time cost of these two approaches. Also we want to see where we are comparing with big integer modular multiplication using GMP library, where we don't use radix representation of the numbers but use the integer type provided by the GMP library.

We gave the same input to the three multiplication functions, and verified the results against each other. Table~\ref{tab:multiplication} shows the time costs of
one multiplication operation using the three different approaches
with regard to $k$ (where $p = r^k + 1$). The time given is in $10^{-6}$ second scale. We can see clearly that the
FFT-based multiplication is faster than the GMP-based one. And the
speedup is more obvious when $k$ increases. But both of our approaches are slower than using pure GMP functions.

\begin{figure}[H]
	\centering
	\includegraphics[totalheight=8cm]{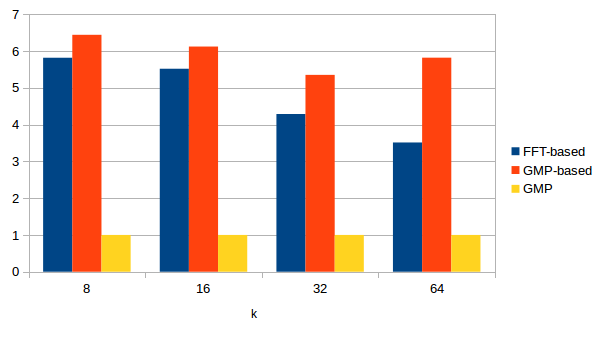}
	\caption{FFT-based multiplication vs. GMP-based multiplication vs. GMP multiplication}
	\label{fig:fftvsgmp}
\end{figure}

Figure \ref{fig:fftvsgmp} shows the cost ratio of FFT-based and GMP-based multiplication versus GMP multiplication. 

\begin{table}[htbp]
	\centering%
	\renewcommand*{\arraystretch}{1.5}
	{
		\begin{scriptsize}
			\begin{tabular}{|l|l|l|l|} \hline 
				$k$ & FFT-based & GMP-based & GMP \\ \hline  \hline
				8&1303.38&	1443.05&	224.03\\ \hline
				16&	2602.56&	2886.63&	471.45\\ \hline
				32&	5500.56&	6865.14	&1282.36\\ \hline
				64&	10656.10&	17649.23&	3032.44\\ \hline
				
			\end{tabular}
		\end{scriptsize}
	}
	\caption{\small Time cost of one multiplication operation using FFT-based, GMP-based and GMP approaches.}
	\label{tab:multiplication}
\end{table}

As introduced in Section~\ref{sec:fft-based implementation}, 
the FFT-based multiplication (in the
generalized Fermat prime field)
takes several steps:
\begin{itemize}
	\item[Step 1] convert the input elements into Montgomery representation 
	\item[Step 2] negacyclic convolution
	\item[Step 3] convert the result out from Montgomery representation 
	\item[Step 4] Chinese Remainder Theorem Algorithm
	\item[Step 5] LHC algorithm
	\item[Step 6] cyclic shift and addition to get the final result.
\end{itemize}

\begin{figure}[H]
	\centering
	\includegraphics[totalheight=8cm]{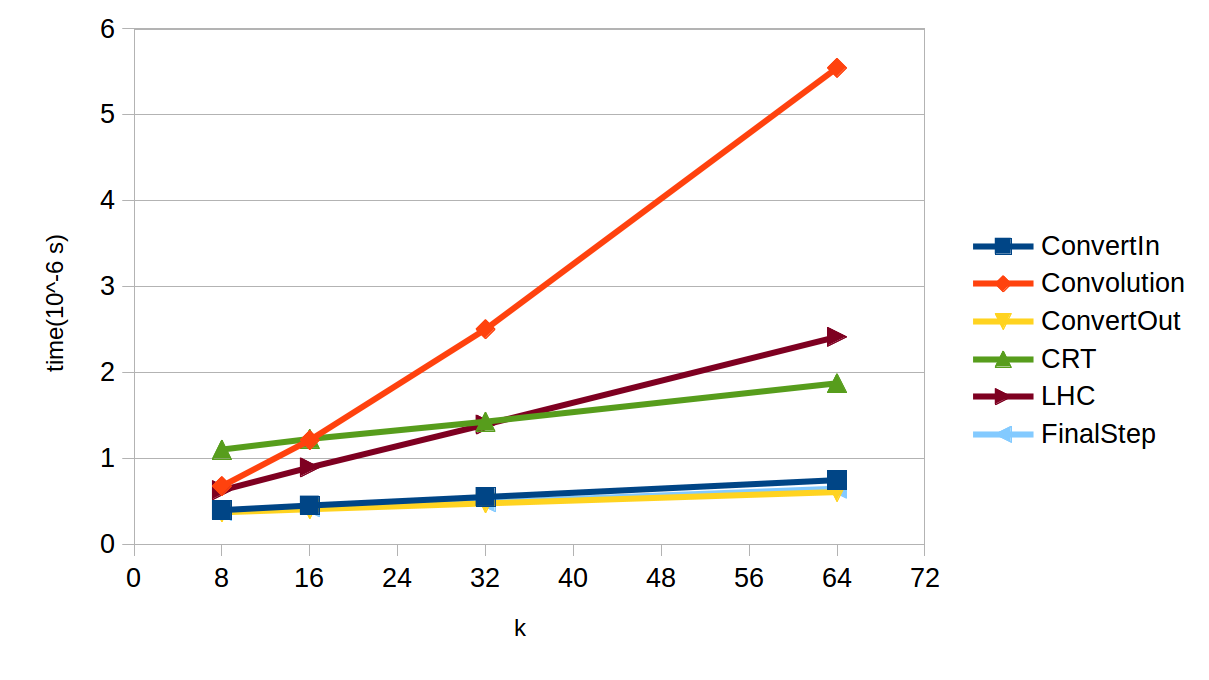}
	\caption{Time spends in different parts of the FFT-based multiplication}
	\label{fig:fftbased}
\end{figure}

Figure~\ref{fig:fftbased} shows the time costs of the above 6 steps $w.r.t$ $k$. 
Table~\ref{tab:fftbasedpercentage} shows the percentage of running 
time for each step over the total time of the multiplication operation. Convolution takes the dominate part of the cost which fits in the analysis we made in Section~\ref{sec:analysisofpolymul}

\begin{table}[htbp]
	\centering%
	\renewcommand*{\arraystretch}{1.35}
	{
		\begin{scriptsize}
			\begin{tabular}{|l|l|l|l|l|l|l|} \hline 
				$k$ & ConvertIn & Convolution & ConvertOut & CRT & LHC & Final step \\ \hline  \hline
				8 & 11.17 & 18.98 & 10.44 & 30.88 & 17.65 & 10.89 \\ \hline
				16 & 9.77 & 26.20 & 8.84 & 26.51 & 19.34 & 9.34 \\ \hline
				32 & 8.06 & 36.59 & 6.97 & 20.86 & 20.37 & 7.16 \\ \hline
				64 & 6.32 & 46.83 & 5.14 & 15.83 & 20.40 & 5.48 \\ \hline		
			\end{tabular}
		\end{scriptsize}
	}
	\caption{\small Time cost in different parts of the FFT-based multiplication in percentage.}
	\label{tab:fftbasedpercentage}
\end{table}

\subsection{FFT over big prime fields}
\label{sec:fftoverbigprimefield}
In this section, we provide experiment data for FFT over big prime fields. 
The FFT function we use is that of  Algorithm~\ref{algo:dft_general}, except that we pre-compute all the power of $\omega$ and passed them as input to 
the algorithm; this is a standard optimization in FFT code
over finite fields~\cite{pan2011algorithmic}.

We compare FFT computation using the arithmetic over the following finite fields:
\begin{itemize}
	\item {\tt GeneralizedFermatPrimeField} in C functions  (FFT-based)\ref{GFPFfftc}
	\item {\tt GeneralizedFermatPrimeField} in C functions (GMP-based)\ref{GFPFgmpc}
	\item {\tt BigPrimeField} in C functions (GMP) \ref{bigprimefieldc}
\end{itemize}
where $K$ is the base-case size and $K^e$ is the input vector size. We should notice that for a prime number $p = r^k + 1$, the base-case size we choose should always satisfy $K = 2k$. Table~\ref{tab:testprimes} gives the prime numbers we use for different base-cases.
\begin{table}[htbp]
	\centering%
	\renewcommand*{\arraystretch}{1.35}
	{
		\begin{scriptsize}
			\begin{tabular}{|l|l|l|l|} \hline
				$K$ &	k & r \\ \hline 
				16 & 8 & $2^{59} + 2^{16}$\\ \hline
				32 & 16 & $2^{58} + 2^{10}$\\ \hline
				64 & 32 & $2^{56} + 2^{21}$\\ \hline
			\end{tabular}
		\end{scriptsize}
	}
	\caption{\small Primes used for different base-cases}
	\label{tab:testprimes}
\end{table}

\begin{figure}[H]
	\centering
	\includegraphics[totalheight=8cm]{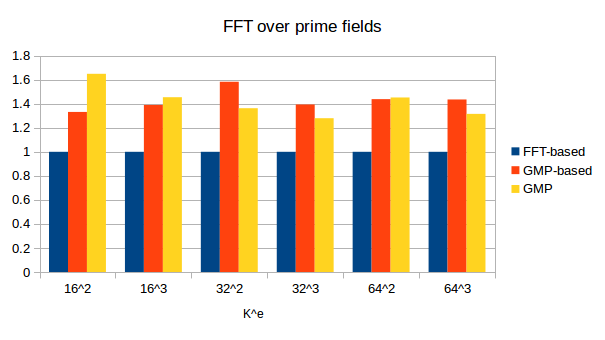}
	\caption{FFT of size $K^e$ where $K = 16$}
	\label{fig:fftoverprimefield}
\end{figure}

Table \ref{tab:fftoverprimefield} gives the time cost of FFT on vector with size $K^e$ over the three prime fields. Figure \ref{fig:fftoverprimefield} shows the cost ratio of GMP-based and GMP versus FFT-based. We can clearly see that FFT over GeneralizedFermatPrimeField using FFT-based multiplication is faster than the other two while BigPrimeField using GMP C functions beats the GMP-based one as the vector size increasing. 

\begin{table}[htbp]
	\centering%
	\renewcommand*{\arraystretch}{1.35}
	{
		\begin{scriptsize}
			\begin{tabular}{|l|l|l|l|l|} \hline 
				$K$ & $e$ & FFT-based & GMP-based & GMP \\ \hline  \hline
				16 & 2 & 0.211	&0.281&	0.348\\ \hline
				16 & 3 &5.961	&8.287&	8.669\\ \hline
				32 & 2 &1.819	&2.49&	2.47\\ \hline
				32 & 3 &109.681	&152.877&	140.342\\ \hline
				64 & 2 &15.775	&22.688	&22.912\\ \hline
				64 & 3 &1995.939	&2865.527&2626.658	\\ \hline
				
			\end{tabular}
		\end{scriptsize}
	}
	\caption{\small Time cost of FFT on vector size $K^e$ over different prime fields}
	\label{tab:fftoverprimefield}
\end{table}

Table~\ref{tab:highprofiling} gives the high-level profiling data on
different steps in the FFT algorithm for $K = 64$ and $e = 3$. For both
FFT-based and GMP-based implementations
of {\tt GeneralizedFermatPrimeField}, most of the time is spent on twiddle
factor multiplication where we need to multiply two arbitrary elements
in the fields. Comparing with the GMP one, we spent less time in the
base-case DFTS, since we only use shift for the multiplication inside the
base-case code and that is where we gain our speed up. This profiling
result agrees with our original thought of using the trick from
F{\"u}rer's paper\cite{DBLP:conf/stoc/Furer07}.

\begin{table}[htbp]
	\centering%
	\renewcommand*{\arraystretch}{1.35}
	{
		\begin{scriptsize}
			\begin{tabular}{|l|l|l|l|} \hline
				time(ms)& permutation& $\mathsf{DFT}_K$ & Twiddle \\ \hline
				FFT-based & 8.08 & 1400.53 & 3460.98 \\ \hline
				GMP-based & 7.84 & 1307.23 & 6996.69 \\ \hline
				GMP & 721.98 & 6418.14 & 1551.41 \\ \hline	
			\end{tabular}
		\end{scriptsize}
	}
	\caption{\small Time spend in different parts of the FFT function when $K = 64,e = 3$}
	\label{tab:highprofiling}
\end{table}

We can see from Figure \ref{fig:fftvsgmp} that for multiplication between two arbitrary elements in a big prime field, the two implementations of ours (FFT-based and GMP-based) are both slower than pure GMP arithmetic. But Figure \ref{fig:fftoverprimefield} shows that for computing a FFT over big vectors, using GeneralizedFermatPrimeField arithmetic with FFT-based multiplication can be more efficient than using pure GMP arithmetic. The main reason is that most of the multiplications are done by the cheap (actually linear time) multiplication, see
Section~\ref{sec:multiplybypowerofr} in the GeneralizedFermatPrimeField while for pure GMP arithmetic all the multiplications are done using the same algorithm. 

\begin{figure}[H]
	\centering
	\includegraphics[totalheight=8cm]{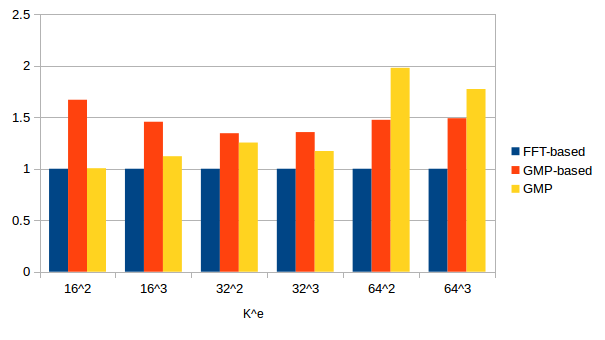}
	\caption{Average time of one multiplication operation in FFT}
	\label{fig:averagetime}
\end{figure}

Table~\ref{tab:averagemultiplication} shows the average time spending in one modular multiplication operation in FFT on vectors with size $K^e$. Figure~\ref{fig:averagetime} gives the radio of GMP-based and GMP versus FFT-based.  We can see that, when computing FFT over Generalized Fermat prime fields, the average time of multiplication operation is less than that of GMP arithmetic. Now we can prove that by using the cheap multiplication with the power of $r$, we can lower the average time spent in multiplication, and further speed up the FFT process. 

\begin{table}[htbp]
	\centering%
	\renewcommand*{\arraystretch}{1.35}
	{
		\begin{scriptsize}
			\begin{tabular}{|l|l|l|l|l|} \hline 
				$K$ & $e$ & FFT-based & GMP-based & GMP \\ \hline  \hline
				16 & 2 &0.000179&0.000299&0.00018\\ \hline
				16 & 3 &0.000197&0.000287&	0.000221\\ \hline
				32 & 2 &0.00031&	0.000417&	0.000389\\ \hline
				32 & 3 &0.000354&	0.00048&	0.000415\\ \hline
				64 & 2 &0.000553&	0.000816&	0.001095\\ \hline
				64 & 3 &0.000652&	0.000972&	0.001157\\ \hline

			\end{tabular}
		\end{scriptsize}
	}
	\caption{\small Average multiplication time of FFT over big prime fields (Time is in ms)}
	\label{tab:averagemultiplication}
\end{table}

\subsection*{Acknowledgements}
The authors would like to thank IBM Canada Ltd (CAS project 880) and
NSERC of Canada (CRD grant CRDPJ500717-16).
{\small
\bibliographystyle{plain}
\bibliography{references}
}
\newpage
	\section*{Appendx A. C Functions for Multiplication in Generalized Fermat Prime Field}
\label{AppA}
\myappendices{Appendix \ref{AppA} }
\begin{lstlisting}[caption={Multiplication between two 128-bit numbers},captionpos=b,label = {lst:u128mulu128}]
//mult_u128_u128_hi128: returns ((x0+x1.u64)*(y0+y1.u64))>>128
void __inline__ mult_u128_u128_hi128(const usfixn64 & x0, const usfixn64 & x1,
const usfixn64 & y0, const usfixn64 & y1, usfixn64 & q)
{
	usfixn64 s0, s1, s2;
	usfixn64 c1;
	
	q = 0;
	
	//	s0 = (__int128) x0 * (__int128) y0;
	// 	s0>>=64;
	__asm__ (
	"movq  %1, %%rax;\n\t"          // rax = a
	"mulq  %2;\n\t"// rdx:rax = a * b
	"movq  %%rdx, %0;\n\t"// s1 = rdx
	: "=rm" (s0)
	: "rm"(x0), "rm"(y0)
	: "%rax", "%rdx");
	
	//	s1 = (__int128) x1 * (__int128) y0;
	//	c1 = (s1 >> 64);
	//	s1 = s1 & (U64_MASK);
	
	__asm__ (
	"movq  %2, %%rax;\n\t"          // rax = a
	"mulq  %3;\n\t"// rdx:rax = a * b
	"movq  %%rax, %0;\n\t"// s1 = rdx
	"movq  %%rdx, %1;\n\t"// s1 = rdx
	: "=rm" (s1),"=rm"(c1)
	: "rm"(x1), "rm"(y0)
	: "%rax", "%rdx");
	
	//	s2 = (__int128) x0 * (__int128) y1;
	//	c2 = (s2 >> 64);
	//	s2 = s2 & (U64_MASK);
	
	__asm__ (
	"movq  %2, %%rax;\n\t"          // rax = a
	"mulq  %3;\n\t"// rdx:rax = a * b
	"movq  %%rax, %0;\n\t"// s1 = rdx
	//			"movq  %%rdx, %1;\n\t"// s1 = rdx
	"addq  %%rdx, %1;\n\t"// s1 = rdx
	: "=rm" (s2),"=rm"(c1)
	: "rm"(x0), "rm"(y1)
	: "%rax", "%rdx");
	
	//	c1+=c2;
	q += c1;
	//	s3 = (__int128) x1 * (__int128) y1;
	q += x1 * y1;
	
	__asm__ (
	"movq  %1, %%rax;\n\t"          // rax = a
	"addq  %2, %%rax;\n\t"// rdx:rax = a * b
	"adcq  $0x0, %0;\n\t"
	"addq  %3, %%rax;\n\t"// rdx:rax = a * b
	"adcq  $0x0, %0;\n\t"
	//			"movq  %%rax, %0;\n\t"// s1 = rdx
	: "+rm"(q)
	: "rm" (s0), "rm"(s1), "rm"(s2)
	: "%rax");
}
\end{lstlisting}

\begin{lstlisting}[caption={Chinese Remainder Algorithm},captionpos=b,label = {lst:crt}]
void crt_mult_sub_u192_with_reduction(const usfixn64 &a1, const usfixn64 &a2,
const crt_u192_data & data, usfixn64 &s0, usfixn64 & s1)
{
	usfixn64 t[4];
	usfixn64 q[2];
	__int128 r[2];
	
	mult_u64_u64(a1, data.m2, t[0], t[1]);
	mult_u64_u64(a2, data.m1, t[2], t[3]);
	
	mult_u128_u128_hi128(t[0], t[1], data.p1_inv_m, data.p1_inv_q, q[0]);
	mult_u128_u128_hi128(t[2], t[3], data.p2_inv_m, data.p2_inv_q, q[1]);
	usfixn64 m0, m1;
	
	__asm__ (
	"movq  %2, %%rax;\n\t"          // rax = a
	"mulq  %3;\n\t"// rdx:rax = a * b
	"movq  %%rax, %0;\n\t"// s1 = rdx
	"movq  %%rdx, %1;\n\t"// s1 = rdx
	: "=rm" (m0),"=rm"(m1)
	: "rm"(q[0]), "rm"(data.p1)
	: "%rax", "%rdx");
	
	m0 = U64_MASK - m0;
	m1 = U64_MASK - m1;
	
	__asm__ (
	"addq %2, %0; \n\t"
	"adcq %3, %1; \n\t"
	"addq $0x1, %0; \n\t"
	"adcq $0x0, %1; \n\t"
	: "+rm" (t[0]),"+rm"(t[1])
	: "rm"(m0), "rm"(m1)
	: );
	
	///////////////////////////////////
	
	m0 = 0;
	m1 = 0;
	
	__asm__ (
	"movq  %2, %%rax;\n\t"          // rax = a
	"mulq  %3;\n\t"// rdx:rax = a * b
	"movq  %%rax, %0;\n\t"// s1 = rdx
	"movq  %%rdx, %1;\n\t"// s1 = rdx
	: "=rm" (m0),"=rm"(m1)
	: "rm"(q[1]), "rm"(data.p2)
	: "%rax", "%rdx");
	
	m0 = U64_MASK - m0;
	m1 = U64_MASK - m1;
	
	__asm__ (
	"addq %2, %0; \n\t"
	"adcq %3, %1; \n\t"
	"addq $0x1, %0; \n\t"
	"adcq $0x0, %1; \n\t"
	: "+rm" (t[2]),"+rm"(t[3])
	: "rm"(m0), "rm"(m1)
	: );
	
	///////////////////////////////////
	
	if (t[0] >= data.p1)
	t[0] -= data.p1;
	if (t[2] >= data.p2)
	t[2] -= data.p2;
	
	//	r[0] = (__int128) t[0] ;///+ ((__int128) t[1] << 64);
	//	r[1] = (__int128) t[2] ;//+ ((__int128) t[3] << 64);
	
	mult_u64_u64(t[0], data.p2, t[0], t[1]);
	mult_u64_u64(t[2], data.p1, t[2], t[3]);
	
	m0 = t[0];
	m1 = t[1];
	__asm__ (
	"addq %2, %0; \n\t"
	"adcq %3, %1; \n\t"
	//					"addq $0x1, %0; \n\t"
	//					"adcq $0x0, %1; \n\t"
	: "+rm" (t[0]),"+rm"(t[1])
	: "rm"(t[2]), "rm"(t[3])
	: );
	if ((t[1] > data.p1p2_q) || ((t[1] == data.p1p2_q) && (t[0] > data.p1p2_m)))
	{
		m0 = U64_MASK - data.p1p2_m;
		m1 = U64_MASK - data.p1p2_q;
		__asm__ (
		"addq %2, %0; \n\t"
		"adcq %3, %1; \n\t"
		"addq $0x1, %0; \n\t"
		"adcq $0x0, %1; \n\t"
		: "+rm" (t[0]),"+rm"(t[1])
		: "rm"(m0), "rm"(m1)
		: );
		
	}
	s0 = t[0];
	s1 = t[1];
}
\end{lstlisting}

\begin{lstlisting}[caption={Computing the quotient and remainder of a machine word size number divided by a radix r},captionpos=b,label = {lst:divbyr}]
void __inline__ div_by_const_R(const usfixn64 x0_u64, const usfixn64 x1_u64,
const usfixn64 r0, const usfixn64 r1, usfixn64 & q, usfixn64 & m)
{
	//	r_inv= (u128/r_in);
	//	r1,r0=[r_inv/u64, r_inv%u64];
	//	x1,x0=[x/u64, x%u64];
	//	v0=x0*r0;
	//	v1=x0*r1;
	//	v2=x1*r0;
	
	__int128 v0, v1, v2, q0;
	usfixn64 x0 = x0_u64;
	usfixn64 x1 = x1_u64;
	v0 = 0;
	v1 = 0;
	v2 = 0;
	
	v0 = (__int128) x0 * (__int128) r0;
	v1 = (__int128) x0 * (__int128) r1;
	v2 = (__int128) x1 * (__int128) r0;
	
	v0 >>= 64;
	v1 += (v0);
	v2 += v1;
	v2 >>= 64;
	
	//	q0 = 0;
	q0 = (__int128) x1 * (__int128) r1;
	q0 += v2;
	
	//	m0=x-(q0*r_in);
	__int128 m0;
	m0 = (__int128) (1L << 64);
	m0 *= (__int128) x1;
	m0 += (__int128) x0;
	
	m0 = m0 - q0 * (__int128) (R);
	
	if (m0 >= (__int128) (R))
	{
		printf("carry\n");
		m0 -= (__int128) R;
		q0 += 1;
	}
	
	m = (usfixn64) (m0 & U64_MASK);
	q = (usfixn64) (q0 & (U64_MASK));
	
	if ((q0 >> 64) > 0)
	{
		printf("WARNING: q >= u64!\n");
	}
}
\end{lstlisting}

\begin{lstlisting}[caption={(l,h,c) Algorithm},captionpos=b,label = {lst:lhc}]
void lhc_by_R_u128(const usfixn64 x0, const usfixn64 x1, const usfixn64 & r0,
const usfixn64 & r1, const usfixn64 & u64_mod_R_q,
const usfixn64 & u64_mod_R_m, usfixn64 & s0, usfixn64 &s1, usfixn64 &s2)
{
	//def div_by_R_u128(x0,x1,x2=0,v=1):
	//
	//	usfixn64 x2 = 0;
	usfixn64 qb, mb;
	usfixn64 m0, q0;
	usfixn64 m1, q1;
	//	### should be precomputed
	//	[mb,qb]=div_by_const_R(u64);
	//	div_by_const_R(0, 1, r0, r1, qb, mb);
	
	qb = u64_mod_R_q;
	mb = u64_mod_R_m;
	//	[m0,q0]=div_by_const_R(x0);
	div_by_const_R(x0, 0, r0, r1, q0, m0);
	
	//	# q1=x1/R;
	//	# m1=x1%R;
	//	[m1,q1]=div_by_const_R(x1);
	
	div_by_const_R(x1, 0, r0, r1, q1, m1);
	
	__int128 l0, l1;
	__int128 h0, h1;
	__int128 c0, c1;
	
	l0 = m0;
	l1 = (__int128) m1 * (__int128) mb;
	//	#l2=0;#x2*R0_u128;
	
	h0 = (__int128) q0;
	h1 = (__int128) q1 * (__int128) mb + (__int128) qb * (__int128) m1;
	//	#h2=0;#x2*R1_u128;
	
	c0 = 0;
	c1 = (__int128) q1 * (__int128) qb;
	//	#c2=0;#x2*R2_u128
	
	
	usfixn64 lhc_l0h0c0[3] =
	{ l0, h0, c0 };
	//	lhc_l0h0c0=[l0,h0,c0];
	
	usfixn64 lhc_l1c1[3] =
	{ 0, 0, c1 };
	usfixn64 lhc_l2c2[3] =
	{ 0, 0, 0 };
	usfixn64 lhc_h1h2[3] =
	{ 0, 0, 0 };
	usfixn64 lhc_ans[3] =
	{ 0, 0, 0 };
	
	div_by_const_R(l1 & U64_MASK, l1 >> 64, r0, r1, lhc_l1c1[1], lhc_l1c1[0]);
	div_by_const_R(h1 & U64_MASK, h1 >> 64, r0, r1, lhc_h1h2[2], lhc_h1h2[1]);
	
	//	lhc_ans=add_lhc(lhc_l0h0c0,lhc_l1c1)
	//	lhc_ans=add_lhc(lhc_ans,lhc_l2c2)
	
	add_lhc(lhc_l0h0c0, lhc_l1c1, lhc_ans);
	add_lhc(lhc_ans, lhc_h1h2, lhc_ans);
	
	//	printf("l=%lu, h=%lu, c=%lu\n", lhc_ans[0], lhc_ans[1], lhc_ans[2]);
	
	s0 = lhc_ans[0];
	s1 = lhc_ans[1];
	s2 = lhc_ans[2];
	
}
\end{lstlisting}

\begin{lstlisting}[caption={Montgomery multiplication for 64-bit numbers},captionpos=b,label = {lst:montmul}]
inline sfixn MontMulModSpe_OPT3_AS_GENE_INLINE(sfixn a,sfixn b,sfixn MY_PRIME, sfixn INV_PRIME){
	asm("mulq %2\n\t"
	"movq %%rax,%%rsi\n\t"
	"movq %%rdx,%%rdi\n\t"
	"imulq %3,%%rax\n\t"
	"mulq %4\n\t"
	"add %%rsi,%%rax\n\t"
	"adc %%rdi,%%rdx\n\t"
	"subq %4,%%rdx\n\t"
	"mov %%rdx,%%rax\n\t"
	"sar $63,%%rax\n\t"
	"andq %4,%%rax\n\t"
	"addq %%rax,%%rdx\n\t"
	: "=d" (a)
	: "a"(a),"rm"(b),"b"((sfixn) INV_PRIME),"c"((sfixn) MY_PRIME)
	:"rsi","rdi");
	return a;
}
\end{lstlisting}


\end{document}